\pdfoutput=1 
\documentclass[useAMS,amsmath,usenatbib,onecolumn]{mn2e}
\usepackage{psfig}  
\usepackage{graphicx} 
\usepackage{dcolumn}  
\usepackage{bm}       
\usepackage{amssymb,amsmath}  
\usepackage{color}
\usepackage{float}
\newcommand{\beq}{\begin{equation}}
\newcommand{\eeq}{\end{equation}}

\newcommand{\beqa}{\begin{eqnarray}}

\newcommand{\etal}{{\it et al.}}

\def\Kpc{\, h^{-1} \, {\rm Kpc}}
\def\Mpc{\, h^{-1} \, {\rm Mpc}}
\def\Gpc{\, h^{-1} \, {\rm Gpc}}

\def\Msun{\,h^{-1}\,{\rm M_{\odot}}}
\def\ltsima{$\; \buildrel < \over \sim \;$}   
\def\gtsima{$\; \buildrel > \over \sim \;$}   
\def\simlt{\lower.5ex\hbox{\ltsima}}   
\def\simgt{\lower.5ex\hbox{\gtsima}}   
\def\etal{{et al. }}

\newcommand{\nc}{\newcommand}   

\nc{\de}{\delta}
\nc{\hn}{\hat{n}}
\nc{\bH}{\bar{H}}
\nc{\Ol}{\Om_{\Lambda}}

\nc{\ul}{\underline} \nc{\al}{\alpha} \nc{\g}{\gamma}
\nc{\Del}{\Delta} \nc{\e}{\textrm{e}} \nc{\eps}{\epsilon}
\nc{\lam}{\lambda} \nc{\Om}{\Omega} \nc{\Omm}{\Omega_m}
\nc{\Oml}{\Omega_\Lambda} \nc{\LCDM}{$\Lambda$CDM~} 
\nc{\ve}{\varepsilon} \nc{\mn}{{\mu\nu}} \nc{\vp}{\varphi}

\def\gsim{\; \raise0.3ex\hbox{$>$\kern-0.75em
\raise-1.1ex\hbox{$\sim$}}\; }

\nc{\Section}[2]{\section{#2}\label{#1}}   
\nc{\Bibitem}[1]{\bibitem{#1}}   
\nc{\Label}[1]{\label{#1}}   

\nc{\hq}{\hat{q}}
\nc{\hw}{\widehat{w}}

\def\ben{\begin{enumerate}}
\def\een{\end{enumerate}}
\def\bi{\begin{itemize}}
\def\ei{\end{itemize}}
\def\ee{\end{equation}}
\def\bea{\begin{eqnarray}}
\def\eea{\end{eqnarray}}

\nc{\M}{\rm{M}}
\nc{\Gpcc}{\rm{~ Gpc^3/h^3}}     
   
\def\etal{{et al. }}   
   
\def\ltsima{$\; \buildrel < \over \sim \;$}   
\def\gtsima{$\; \buildrel > \over \sim \;$}   
\def\simlt{\lower.5ex\hbox{\ltsima}}   
\def\simgt{\lower.5ex\hbox{\gtsima}}   
\nc{\w}{$w(\theta)$\ }   
\nc{\ie}{i.e., }    
\nc{\eg}{e.g., }

\input epsf

\title[The MICE Grand Challenge]
{The MICE Grand Challenge Lightcone Simulation III:\\ 
Galaxy lensing mocks from all-sky lensing maps}

\author[Fosalba \etal]{
P. Fosalba,  E. Gazta\~{n}aga,  F. J. Castander  \& M. Crocce\\
Institut de Ci\`encies de l'Espai, IEEC-CSIC, Campus UAB, Facultat de
Ci\`encies, Torre C5 par-2, Barcelona 08193, Spain}

\begin{document}

\twocolumn   
\maketitle 

\begin{abstract}
In paper I of this series (Fosalba et al. 2015), we 
presented a new N-body lightcone simulation from the MICE collaboration, {\it the
MICE Grand Challenge} (MICE-GC), containing about 70 billion
dark-matter particles in a $(3 \Gpc)^3$ comoving volume, from which we built
halo and galaxy catalogues using a Halo Occupation Distribution and
Halo Abundance Matching technique, as presented in the companion Paper II (Crocce et
al. 2015). Given its large volume and fine mass resolution,
the MICE-GC simulation also allows an accurate modeling of the lensing
observables from upcoming wide and deep galaxy surveys.
In the last paper of this series (Paper III), we describe the construction
of all-sky lensing maps, following the ``Onion Universe'' approach (Fosalba et
al. 2008), and discuss their properties in the lightcone up to
$z=1.4$ with sub-arcmin spatial resolution.
By comparing the convergence power spectrum in the MICE-GC to
lower mass-resolution (\ie particle mass $\sim 10^{11} \Msun$) simulations, 
we find that resolution effects are at the 5$\%$ level for multipoles $\ell\sim 10^3$
and 20 $\%$ for $\ell\sim 10^4$. Resolution effects have a much lower
impact on our simulation, as shown by comparing the MICE-GC to recent
numerical fits by Takahashi et al 2012.
We use the all-sky lensing maps to model
galaxy lensing properties, such as the
convergence, shear, and lensed magnitudes and positions, and
validate them thoroughly using galaxy shear auto and cross-correlations in harmonic and
configuration space.
Our results show that the galaxy lensing mocks here presented can be used
to accurately model lensing observables down to arc-minute scales.
Accompanying this series of papers, we make a first public data
release of the MICE-GC galaxy mock, the {\tt MICECAT v1.0}, through a dedicated
web-portal for the MICE simulations:
{\texttt http://cosmohub.pic.es},
to help developing and exploiting the new generation of astronomical surveys.

\end{abstract}   
   

\section{Introduction}   

Thanks to the new generation of large astronomical surveys, we have
entered in the era of precision cosmology. The high quality data 
that will be collected by upcoming galaxy surveys, such as 
DES\footnote{\texttt{www.darkenergysurvey.org}}, 
HSC\footnote{\texttt{www.naoj.org/Projects/HSC}}, 
Euclid\footnote{\texttt{www.euclid-ec.org}}, 
DESI\footnote{\texttt{desi.lbl.gov}},
HETDEX\footnote{\texttt{hetdex.org}}, 
LSST\footnote{\texttt{www.lsst.org}}, 
WFIRST\footnote{\texttt{wfirst.gsfc.nasa.gov}},etc.
will allow to characterize in
great detail the distribution of galaxies from the largest accessible
scales where linear theory applies down to very small scales described by the non-linear regime
of gravitational clustering. Combining observables to beat down
probe-specific systematics has the potential to maximize the
scientific return of these surveys \citep{albrecht06,albrecht07}. In particular, traditional probes such as galaxy
clustering and cluster abundance can be uniquely complemented with
weak lensing data to break degeneracies in cosmological parameters
\citep{weinberg13}.
State-of-the-art weak-lensing observational results have been obtained
by the CFHTLenS survey\footnote{www.cfhtlens.org} (see \cite{kilbinger13} and references
therein), and build upon previous observations over the last to
decades, as summarized in recent reviews \citep{bartelmann01,waerbeke03,hoekstra08,bartelmann10}.

The ultimate goal of this new generation of surveys is to
constrain the nature of dark-energy as well as pin down possible
deviations from the standard model, described by General Relativity
\citep{weinberg13}.
But in order to achieve these goals, it is critical to match such experimental efforts 
with theoretical ones to help developing the science
case of these surveys as well as optimally exploiting
these large and complex observational datasets.

Theoretical modeling of weak-lensing observables is challenging
because the lensing correlations on sub-degree angular scales are
described by the non-linear regime of gravitational clustering, where
a purely analytic description is not possible. In the absence of a compelling analytic description,
Nbody numerical simulations can be used to accurately model the growth of large-scale
structures in the non-linear regime and the lensing distortions they
produce.  Numerical simulations of the weak gravitational lensing
are typically based on ray-tracing techniques through Nbody
simulations \citep{blandford91,wambsganss98,jain00,white00,hamana02,vale03,white04,hilbert09,becker13}.
In the ray-tracing approach, light rays are back-traced from the
observer to the source, as they are deflected from multiple (typically
few tens) of projected-mass lens planes. 
Measurements of the lensing second and higher-order moments in ray-tracing simulations have
been shown to be in good agreement with non-linear theory predictions of
gravitational clustering (see \eg \citep{gaztanaga98,waerbeke01}).
 
However, in order to model lensing observables accurately one needs to cover a
wide dynamical range: from the large linear (few degree) scales 
where the power of the deflection field peaks, down
to the small (few arcmin) scales that capture
the non-linear growth of structures and their associated non-Gaussian
contribution to the lensing covariances \citep{semboloni07,sato09,harnois12}.
On the other hand, modeling lensing observables with ray-tracing
techniques over a significant portion of the sky
are prohibitive in terms of CPU time and memory requirements, and
this method is typically restricted to small (few sq.deg.) patches of the sky.

Alternative methods, where the Nbody matter is projected
along unperturbed paths using the single-plane (or Born) approximation,
have been successfully implemented over large-volume
high-resolution simulations to model weak gravitational lensing on
curved skies \citep{gaztanaga98,fosalba08,das08,teyssier09}. This novel
technique can be readily used to accurately model weak-lensing in
wide field galaxy surveys (for sources at $z_s \simlt 3$) or the CMB
lensing (for $z_s \approx 1100$).\footnote{In fact \cite{das08} generalize
  this approach to the multiple-plane case, in the context of CMB
  lensing, to go beyond the Born approximation}

In Paper I of this series \citep{paperI} we presented a new N-body simulation developed by the MICE
collaboration at the Marenostrum supercomputer, the MICE {\it Grand
Challenge} run (MICE-GC), that includes about 70 billion dark-matter
particles, in a box of about $3 \Gpc$ aside. This simulation samples 5 orders of
magnitude in dynamical range, covering from the largest (linear) scales
accesible to the observable universe where clustering statistics are Gaussian, down to 
to the highly non-linear regime of structure formation where gravity
drives dark-matter and galaxy clustering away from Gaussianity.
Using the MICE-GC dark-matter outputs in the lightcone, 
we built a mock galaxy catalog using a hybrid Halo Occupation
Distribution and Halo Abundance Matching (HOD+HAM) technique, whose galaxy
clustering properties were discussed in detail in Paper II \citep{paperII}.

In the last paper of the series (Paper III)
we describe the construction of all-sky lensing maps, with sub-arcmin
resolution and discuss their application to model galaxy lensing properties, such as the
convergence, shear, and lensed magnitudes and positions of
the synthetic HOD+HAM galaxy catalog. 

Following up on the analyses presented in the previous papers of this series,
one of the main focus of this work is to 
investigate the  impact of {\it mass-resolution  effects} 
in the modeling of dark-matter
and galaxy clustering observables by comparing the MICE-GC, and
previous MICE runs, to analytic fits available based on
high-resolution N-body simulations.
For this purpose, we have used the ``Onion Universe'' approach
\citep{fosalba08}
to build lensing observables from pixelized 2D maps of the dark-matter
in the lightcone. Starting from the convergence, we derive other
observables including the shear, deflection and lensed positions and magnitudes.
We have then developed several tests to validate 
this observables using basic lensing statistics, such as the convergence and shear
angular power spectrum and the shear 2-point
correlation functions, and the cross-correlations of foreground and
background galaxy samples to extract the magnification signal.
Our analysis shows that the all-sky lensing maps and galaxy lensing
properties derived from them can be used to model upcoming galaxy
surveys with high accuracy from the largest (linear) scales down to
the small ($\sim$ arcmin) scales described by the non-linear
regime of gravitational clustering.

This paper is organized as follows: 
\S\ref{sec:sim} briefly describes the MICE-GC
run and its parameters.
In \S\ref{sec:lens} we describe the construction of all-sky
maps of lensing observable such as convergence, shear and the deflection
field, from the dark-matter outputs in the lightcone, and in
\S\ref{sec:gallens} we show how we assign lensing properties to the 
mock galaxies, and validate our implementation by using 2-point shear
auto and cross-correlation statistics in harmonic and configuration space.
We also present a novel application of our all-sky lensing maps: the
modeling of magnification in mock galaxy positions and magnitudes is discussed in \S\ref{sec:magbias}
Finally, in \S\ref{sec:conclusions} we summarize our main results and conclusions.

\section{The MICE Grand Challenge Simulation}   
\label{sec:sim}

In Paper I of this series \citep{paperI},
we presented and validated a new large volume cosmological simulation, 
the MICE \footnote{further details about the MICE project can be found here \texttt{ www.ice.cat/mice}} {\it
  Grand Challenge} simulation (MICE-GC hereafter), developed at the
Marenostrum supercomputer at 
BSC \footnote{Barcelona  Supercomputing Center, \texttt{ www.bsc.es}},
using the public Nbody code Gadget2 \citep{springel05}.
The MICE-GC simulation contains $4096^3$ dark-matter particles in a box-size of $L_{box}=
3072 \Mpc$ (\ie samples a cosmological volume of $\sim 30 \Gpc^3$),
and the softening length used is, $l_{soft} =  50 \rm{kpc/h}$.
Therefore this simulation covers a very wide dynamic range, close to five orders
of magnitude in scale, with a good mass resolution, $m_p = 2.93 \times 10^{10} \Msun$.

In Table \ref{simtab} we describe the Gadget-2 code parameters used in the MICE
simulations discussed in this paper: the MICE Grand-Challenge (MICE-GC),
the Intermediate Resolution (MICE-IR; \cite{fosalba08}) and the Super-Hubble
Volume (MICE-SHV; \cite{crocce10}).
Further details about the MICE-GC run and the validation of its
dark-matter outputs using various 3D and 2D clustering statistics are
given in Paper I \citep{paperI}.

\begin{table*} 
\begin{center}
\begin{tabular}{lcccccccccccccccc}
\hline \\
Run        &&&    $N_{{\rm part}}$ & \ $L_{{\rm box}}/\Mpc$  \ & \ $PMGrid$ &
$m_p/(10^{10} \Msun)$  & $l_{{\rm soft}}/\Kpc$ & $z_{{\rm i}}$   & $Max. TimeStep$ \\

           &&&  &  & &  & & &    \\

MICE-GC   &&&    $4096^3$  & $3072$    & $4096$    & $2.93$  & $50$  & $100$ & $0.02$       \\

           &&&  &  & &  & & &    \\

MICE-IR   &&&    $2048^3$  & $3072$     & $2048$   & $ 23.42$ &  $50$ & $50$   & $0.01$    \\ 

MICE-SHV   &&&    $2048^3$  & $7680$    & $2048$    & $366$   & $50$ & $150$   & $0.03$ \\  \\

\hline
\end{tabular}
\end{center}
\caption{Description of the MICE N-body simulations. $N_{{\rm part}}$
  denotes number of particles, $L_{{\rm box}}$ is the box-size, $PM
  Grid$ gives the size of the Particle-Mesh grid used for the
  large-scale forces computed with FFTs, $m_p$ gives the particle mass, $l_{soft}$ is the softening length,
and  $z_{in}$ is the initial redshift of the simulation. All
simulations had initial conditions generated using the Zeldovich
Approximation. Max. Timestep is the initial global
time-stepping used, which is of order $1\%$
of the Hubble time (i.e, $d \log a=0.01$, being $a$ the scale factor).
The number of global time-steps to complete the runs were $N_{steps}
\simgt 2000$ in all cases, except for the MICE-GC which took 
about $ 3000$ time-steps.
Their cosmological parameters were kept constant throughout the runs
(see text for details).} 
\label{simtab}
\end{table*}


\section{All-sky Lensing maps}
\label{sec:lens}

Following the approach presented in \cite{fosalba08}, 
construct a lightcone simulation by replicating the simulation
box (and translating it) around the observer.  
Given the large box-size used for the MICE-GC simulation,
$L_{box}=3072 \Mpc$,  this approach allows us to build
all-sky lensing outputs with negligible repetition up to $z_{max}=1.4$.
Then we decompose the dark-matter lightcone into a set of all-sky concentric spherical shells,
of given width $\Delta_r$, around the observer, what we call the ``onion
universe''. Each dark-matter ``onion shell'' is then projected onto a 2D
pixelized map using the Healpix tessellation\footnote{{\texttt
    http://sourceforge.net/projects/healpix}}\citep{gorski05}.  
For the lensing maps presented in this paper we have chosen a shell-width of $\Delta_r \approx 35$
megayears in look-back time, which corresponds to a redshift width, $dz\approx 0.003(1+z)$, 
and an angular resolution of $\Delta_{\theta} \approx \sqrt{3/\pi} \, 3600/$
Nside  $\approx 0.85$ arcmin, for the Healpix map resolution
Nside $=4096$ that we use. 
We note that this angular resolution is always larger than the angle
$\theta_{soft}$ subtended by the softening length
used for the MICE-GC run, $l_{soft} =  50 \rm{kpc/h}$, at the redshifts of interest to
derive our lensing observables. In particular, even for the lowest
redshift sources, say at $z=0.2$, $\theta_{soft}=0.36$ arcmin, and
therefore the convolution of the softening length with top-hat
pixel window function smooths the gravity force up to angular scales
$\theta_{smooth}\simeq 2.3 \times \theta_{soft}=0.83$ arcmin, within the pixel
scale used\footnote{Note that if we use maps with the Healpix
  resolution Nside$=8192$, that correspond to a pixel scale of $\Delta_{\theta}
  \approx 0.43$ arcmin, we expect the gravitational softening
  used in the MICE-GC simulation to affect up to
$\theta_{smooth} \simeq 2.3 \times \theta_{soft}  = 0.43$ arcmin 
at z=0.5, and therefore the lensing observables will be 
affected by the softening length effects for lens contributions at redshifts, $z<0.5$}.

Using these redshift and angular resolutions, we can decompose the lightcone
volume in the range $0 < z <1.4$  into 265 onion shells, each containing
$n_{pix}=12\times$ Nside$^2=201.326.592$ pixels.
With this set of pixelized lensing maps we construct a finely gridded
lightcone output, containing $n_{pix}\times$ (number of
z-bins)$=53.351.546.880$ pixels (i.e, $53+$ 3D Gigapixels) with
comoving volume,  $\Delta_V=\Delta_{\theta}^2 r^2 \Delta_r$. 
Total pixelized data volume, using single precision, is
about $200$ GB. This represents a data compression factor of $50$ with
respect to the original $10+$ TB of lightcone output, and thus allows
for a much more efficient post-processing analysis. 

By combining the dark-matter ``onion shells''  that make up the
lightcone, we can easily derive lensing observables, as explained in
\cite{fosalba08}.  This approach, based on approximating the
observables by a discrete sum of  2D dark-matter density maps
multiplied by the appropriate lensing weights, agrees with the much more
complex and CPU time consuming ray-tracing technique
within the Born approximation, i.e., in the limit where lensing deflections are calculated
using unperturbed light paths

\begin{figure}
\begin{center}
\includegraphics[width=0.48\textwidth]{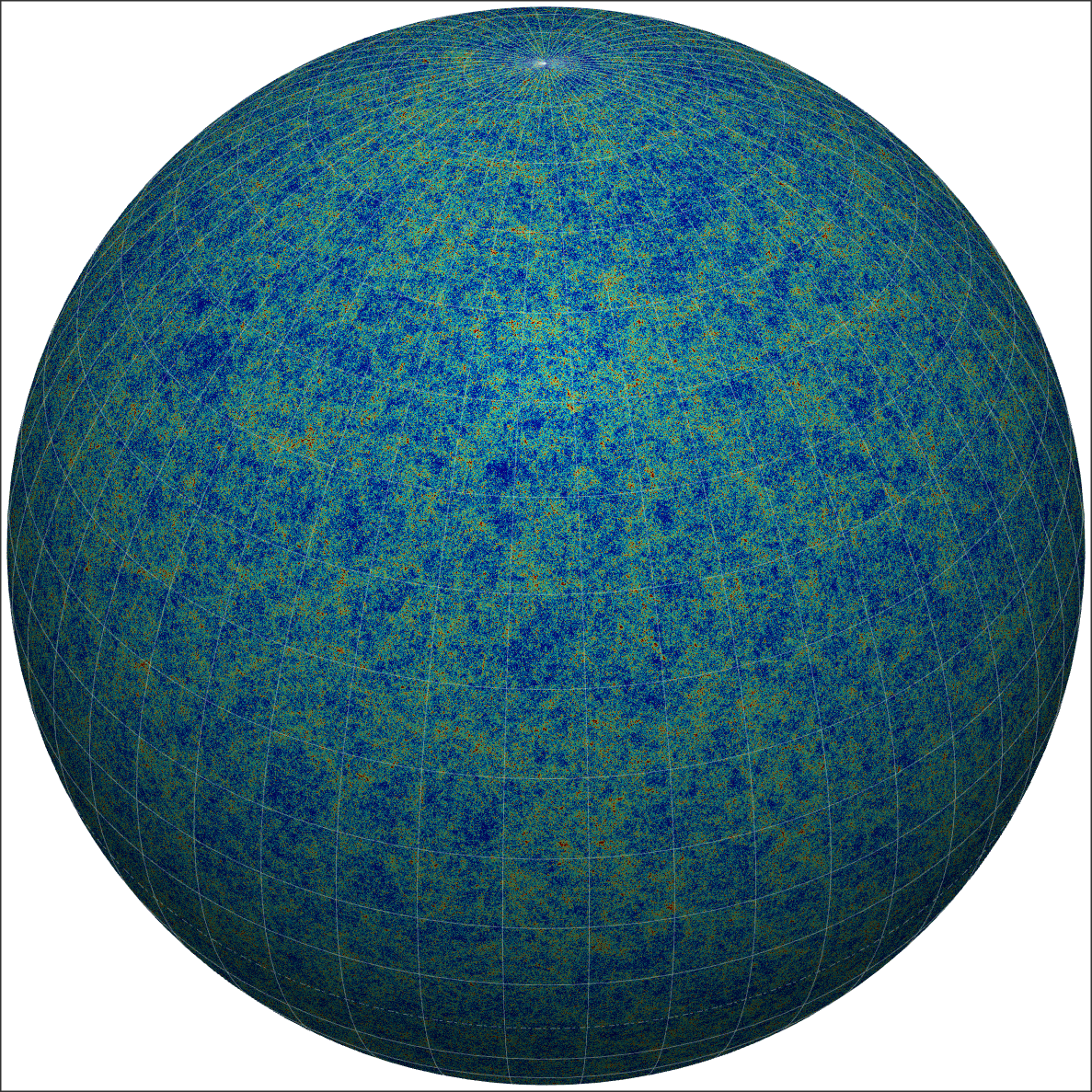}
\caption{
All-sky map of the convergence field, $\kappa$, for the MICE-GC simulation, 
for sources at $z_s=1$, with a pixel resolution of $0.85$ arcmin. 
The sphere is gridded in $\sim 50$ sq.deg patches. The color scale shown spans over the range
$-\sigma < \kappa< 3\, \sigma$, where $\sigma$ is the rms fluctuation
of the all-sky convergence map, illustrating the richness in the
lensing structures resolved.}
\label{fig:mgckappa}
\end{center}
\end{figure}

Following this technique we are able to produce all-sky maps of the convergence field, as well
as maps for other lensing fields obtained from covariant derivatives of the
lensing potential, such as the deflection angle, convergence, shear, flexion, etc. 
Figure~\ref{fig:mgckappa} shows the all-sky map of the convergence
field, $\kappa$, for the MICE-GC simulation, for sources at $z_s=1$, with a pixel resolution of $0.85$ arcmin (\ie
Healpix resolution parameter Nside=4096), produced with the CMBview
software\footnote{\texttt{ https://code.google.com/p/cmbview}}. 
The sphere is gridded in $50$ sq.deg patches. The color scale shown spans over the range
$-\sigma < \kappa< 3\, \sigma$, where $\sigma$ is the rms fluctuation
of the all-sky convergence map, illustrating the richness in the
lensing structures resolved. 

In the all-sky limit, we can take the spherical transform of the
lensing potential to obtain other lensing observables through simple
relations (see \cite{hu00}). In the same way it happens with Fourier transforms in the
flat-sky limit, spatial derivatives in real space translate into
simple multiplications by the corresponding wavenumber (or multipole
for curved sky) in the transformed space, as explained in
more detail below.

\subsection{Convergence}

In what follows we shall compute the convergence field for lensing distortions over
unperturbed paths, the so-called Born approximation,
\beq
\kappa(\theta) = {3 H_0^2\Omega_m\over{2c^2}}~\int ~dr~\delta(r,\theta)
{(r_s-r)r\over{r_s~a}} 
\label{eq:kappa}
\eeq 
where $H_0 = 100 h$ km/s/Mpc is the Hubble constant, $c$ is the speed of light,  $\delta$ is the 3D matter overdensity at
radial distance $r(a)$ (for a corresponding scale factor $a$), and angular position $\theta$, and $r_s$ is the
distance to the lensing sources\footnote{We have assumed flat space, to
be consistent with the cosmology used for the MICE-GC run, but this
can be trivially generalized for non-flat spaces}.
Using the ``Onion Universe'' approach, we can
build a pixelized 2D map of the convergence field in the Born
approximation by simply adding up the dark-matter
``onion shells'' or projected density maps in the lightcone, 
weighted by the weak-lensing efficiency at each redshift,
\beq
\kappa_i = {3H_0^2\Omega_m\over{2c^2}}~\sum_j ~\delta_{i,j}~{(r_s-r_j)r_j\over{r_s a_j}}~dr_j 
\label{eq:kappasum}
\eeq
where $i$ indicates a pixel position in the sky and $j$ a radial bin
(at distance $r_j$ of width $dr_j$) into which we have sliced 
the simulation (for further details see \cite{fosalba08}).

Using Eq.(\ref{eq:kappa}) one gets for the angular
power spectrum of the convergence field in the Born approximation,
\beq
C_{\ell}(\kappa) =
{9H_0^4\Omega_m^2\over{4c^4}}~\int ~dr~ P(k,z) 
{(r_s-r)^2\over{r_s^2~a^2}}
\label{eq:clkappa}
\eeq
where $P(k,z)$ is the 3D density power spectrum
in the simulation at redshift $z$ (corresponding to the radial
coordinate $r=r(z)$ in the integral). For the non-linear theoretical predictions, we
shall replace $P(k,z)$ above with the numerical fits, such as the Halo
model fit to numerical simulations, Halofit
\citep{smith03}, or the revised Halofit \citep{takahashi12}, as
implemented in {\tt CAMB sources} code\footnote{\texttt{http://camb.info/sources/}}.

In turn,
the convergence field, $\kappa$, is related to the lensing potential through the
2D equivalent to the usual (3D) Poisson equation,
\beq
\kappa(\hat{ \rm{n}}) = \nabla^2 \phi(\hat{\rm{n}}) 
\eeq
where $\phi (\hat{\rm{n}})$ is the lensing potential at a given point on
the celestial sphere, denoted by $\hat{\rm{n}}$. 
The coefficients of the spherical harmonic transform,
$\kappa(\hat{\rm{n}})=\sum_{\ell,m} \kappa_{\ell m} Y_{\ell m}(\hat{\rm{n}})$ are given by,
\beq
\kappa_{\ell m} = -\frac{1}{2}\ell(\ell+1) \phi_{\ell m}
\eeq
One can thus use this expression to derive the lensing potential
at each source plane (or 2D lightcone map), and obtain other
lensing observables through their relation to the lensing potential in
harmonic space.

\begin{figure}
\includegraphics[width=0.48\textwidth]{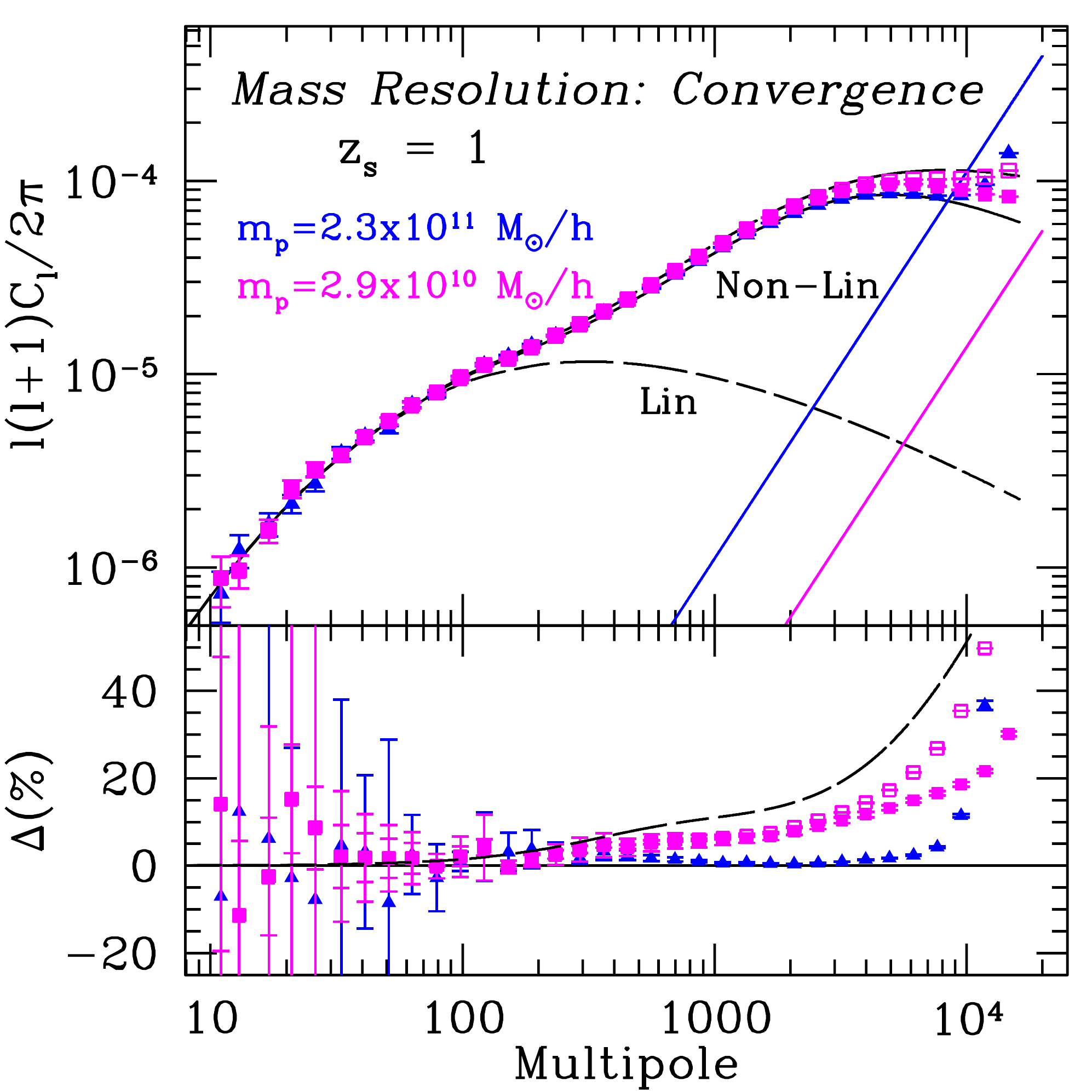} 
\caption{Angular power spectrum of the Convergence map for sources at
 $z_s=1$. Dashed, solid and long-dashed lines show linear theory,
 the Halofit (Smith et al 2003) and revised Halofit (Takahashi et al
 2012) non-linear theory predictions, respectively. Lower panel shows relative
 deviations with respect to theory predictions. Filled symbols show
 simulations measurements including shot-noise, open symbols display
 measurements without shot-noise (shown only for the MICE-GC, where
 its is a subdominant effect). We show Gaussian error bars to guide the
 eye (see text for details). Mass resolution effects estimated
 as the difference between the MICE-IR and the MICE-GC simulations, are at the 5$\%$ level for $\ell\sim 10^3$
and 20 $\%$ for $\ell\sim 10^4$. In turn that MICE-GC displays a
comparable lack of power with respect to the revised Halofit prediction.}
\label{fig:clkmassres}
\end{figure}

Figure~\ref{fig:clkmassres} shows the angular power spectrum of the convergence
map, $C_{\ell}^{\kappa}$, or more precisely,
$\Delta^2_{\kappa}(\ell) \equiv l(l+1) C_{\ell}^{\kappa}/2\pi$, the contribution to the convergence field
variance, $k^2_{rms}$, per logarithmic interval in wavenumber $\ell$, 
for sources at  $z_s=1$. The agreement between
the Halofit predictions and simulations is within $10\%$ for MICE-IR, and $20\%$ for
MICE-GC, down to the resolution scale of the maps used, which
corresponds to a multipole $\ell = 2 \times$ Nside $\sim
16000$\footnote{We use the Healpix {\tt anafast} routine, with ring weights, to compute
  spherical transforms, which are in principle accurate within few
  percent up to $\ell_{max}  \simeq 3$ Nside (see
  \texttt{http://healpix.sourceforge.net/html/facilities.htm}. 
However we take a  conservative approach and only include multipoles up to $\ell_{max}$=
  2 Nside in our analysis.}. 
The excess of power we find in the MICE-GC convergence maps with
respect to the Halofit is in quantitative agreement with a
similar analysis (although limited to a small patch of the sky)
performed over the 
Millennium Simulation (see Fig.9
in \cite{hilbert09}), which has more than one order of magnitude lower
particle mass (\ie $\sim 10^{9} M_{\sun}/h$).

Comparing the power measured in MICE-GC to that of MICE-IR, we conclude that mass 
resolution effects are at the 5$\%$ level for $\ell\sim 10^3$
and 20$\%$ for $\ell\sim 10^4$, what is consistent with what was
observed for the dark-matter clustering in 2D at the
peak of the weak lensing efficiency, \ie $z=0.5$ (see Fig.7 of Paper
I,\citep{paperI}). 
In turn, the recent ``revised'' Halofit
prediction \citep{takahashi12}, based on suite of smaller box
realizations, shows a comparable power excess with
respect to MICE-GC, what seems to indicate that our simulation
might still suffer from mass resolution effects. However, the
discrepancy found could be partially due to 
the difference in box size used between our simulation and the
higher-resolution runs by \cite{takahashi12}. A
more complete (and consistent) analysis
of mass resolution effects in lensing observables and its possible correlation with 
other simulation parameters is left for future work.

\subsection{Deflection Angle}

\begin{figure}
\includegraphics[width=0.48\textwidth]{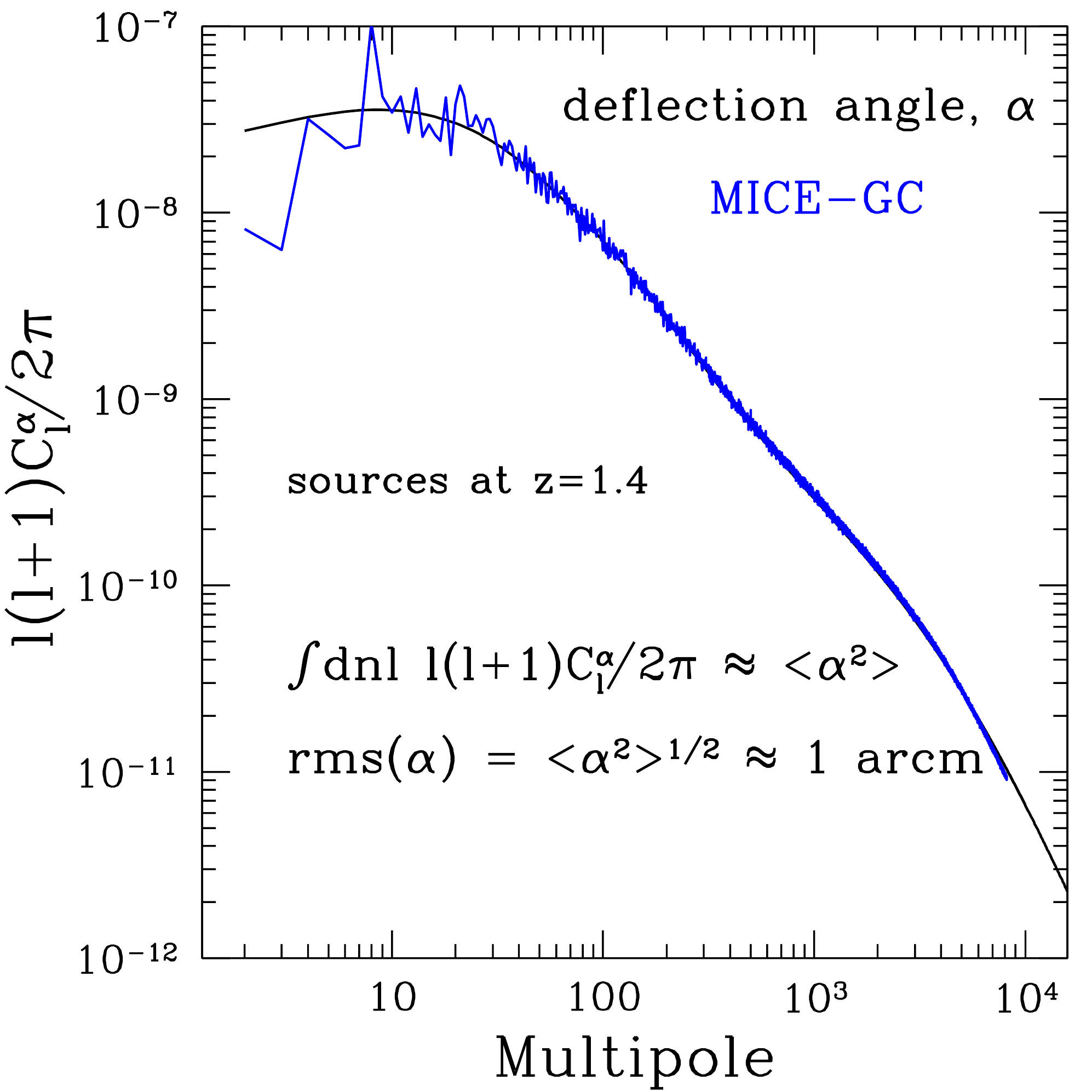} 
\caption{Angular power spectrum of the deflection angle for sources at
  $z_s=1.4$ measured in  the MICE-GC simulation (wiggly blue line), compared to the
 non-linear Halofit prediction (smooth black line).} 
\label{fig:cldefl}
\end{figure}

The gradient of the lensing potential gives the deflection angle, \cite{hu00},
\beq
\alpha(\hat{\rm{n}}) = \nabla \phi(\hat{\rm{n}}) 
\eeq
and the coefficients of its spherical harmonic transform read,
\beq
\alpha_{\ell m} = -\sqrt{\ell(\ell+1)} \phi_{\ell m}
\eeq
so that the corresponding power spectra are simply related, $C_{\ell}^{\alpha}= \ell(\ell+1) C_{\ell}^{\phi}$.
Figure~\ref{fig:cldefl} shows a comparison between the power
spectrum measured in the MICE simulation and the non-linear
theoretical fit (\ie Halofit, see solid line), for sources at 
 $z_s=1.4$. Simulation power spectrum agrees very well with Halofit at
 all scales. At the lowest multipoles sample variance introduces large
 fluctuations in the measured power for a single realization.
The square root of the integral under the curve gives the rms
fluctuation of the deflection field, 
$<\alpha^2>^{1/2} \approx 1$ arcmin.

\subsection{Shear}

In turn shear maps, $\gamma (\hat{\rm{n}})$, can
be simply obtained by decomposing all-sky lensing maps in spherical
harmonics (see \cite{hu00}):
\beq
  \label{eq:shear}
\gamma_{\ell m} = - f(\ell) \kappa_{\ell m}  = \frac{1}{2} f(\ell)
\ell(\ell+1) \phi_{\ell m} ,   
\eeq
with, $f(\ell)=\sqrt{(\ell+2)(\ell-1)/(\ell(\ell+1))}$.
Assuming that, for the cosmological weak-lensing signal, the B-mode is zero, 
the shear E-mode harmonic coefficients read,
$E_{\ell m}=\gamma_{\ell m}$, whereas the $(\gamma_1,\gamma_2$)
"Stokes " parameters of the shear field,
\beq
\gamma_1(\hat{\rm{n}}) \pm i\gamma_2(\hat{\rm{n}}) = \sum_{\ell m}
\gamma_{\ell m} Y_{\ell m}(\hat{\rm{n}})
\eeq
are then obtained transforming back the $E_{\ell m}$'s to real space.

\begin{figure}
\begin{center}
\includegraphics[width=0.48\textwidth]{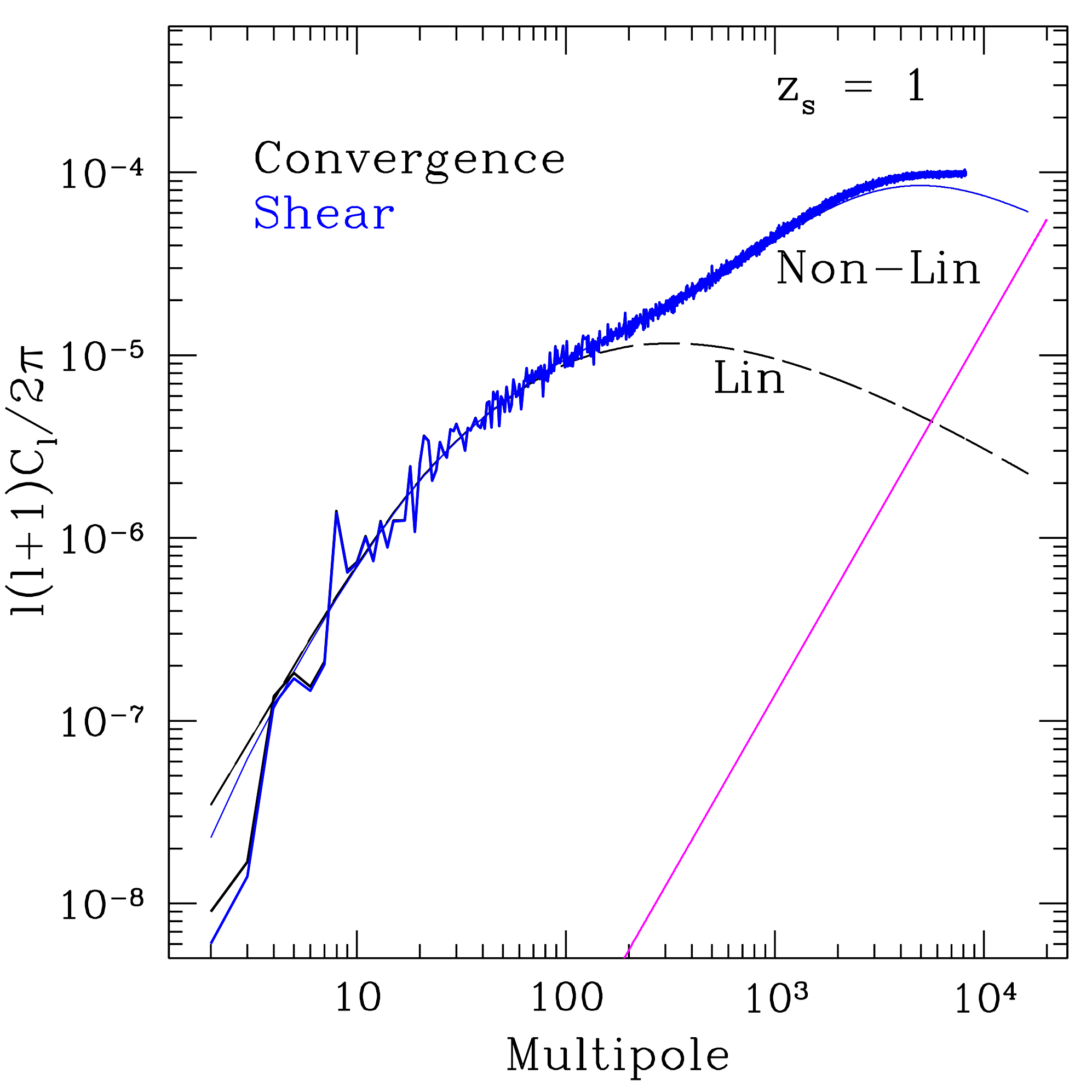}
\caption{Comparison between simulation measurements (wiggly thick solid
  lines) and non-linear theory expectations (Halofit, smooth thin lines) for the convergence (black) and shear
  amplitudes (blue), for sources at $z_s=1$. Shot-noise (magenta
  line) is taken into account and subtracted from the power measured
  in the simulation.}
\label{fig:clshear}
\end{center}
\end{figure}

Fig.\ref{fig:clshear} shows the angular power spectrum $C_{\ell}'s$ for
the convergence (black lines) and shear amplitude (blue) for
dark-matter at source redshift, $z_s=1$ in the
MICE-GC simulation (wiggly lines), as compared to non-linear theory
predictions given by Halofit (smooth lines).  The shot-noise
contribution (bottom magenta straight line) has been subtracted 
from the power measured in the simulation.
Very good agreement is found between non-linear theory and simulations for
a wide dynamical range, similar to what was found for the deflection angle.

\begin{figure*}
\begin{center}
{\includegraphics[width=0.45\textwidth]{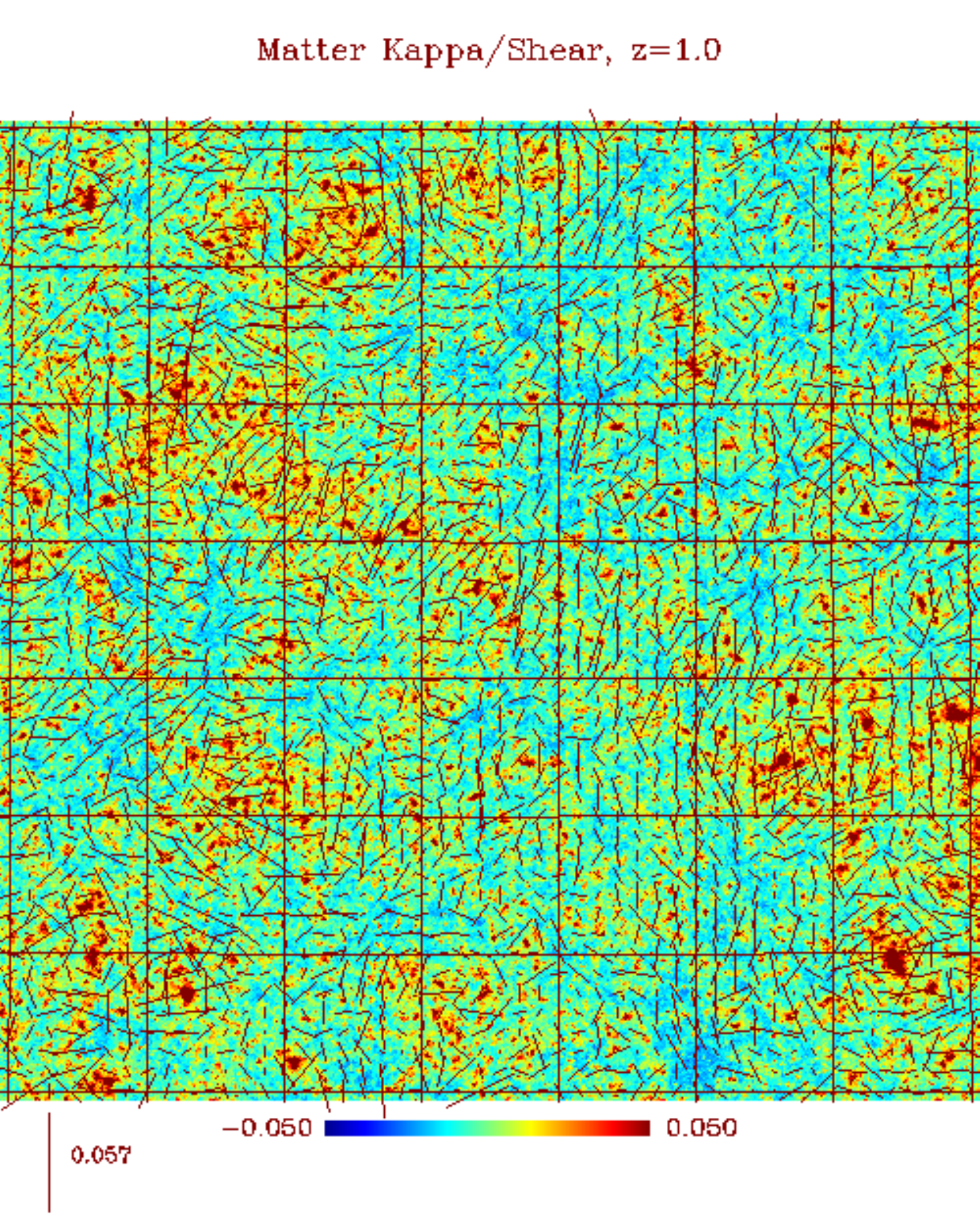}
\includegraphics[width=0.45\textwidth]{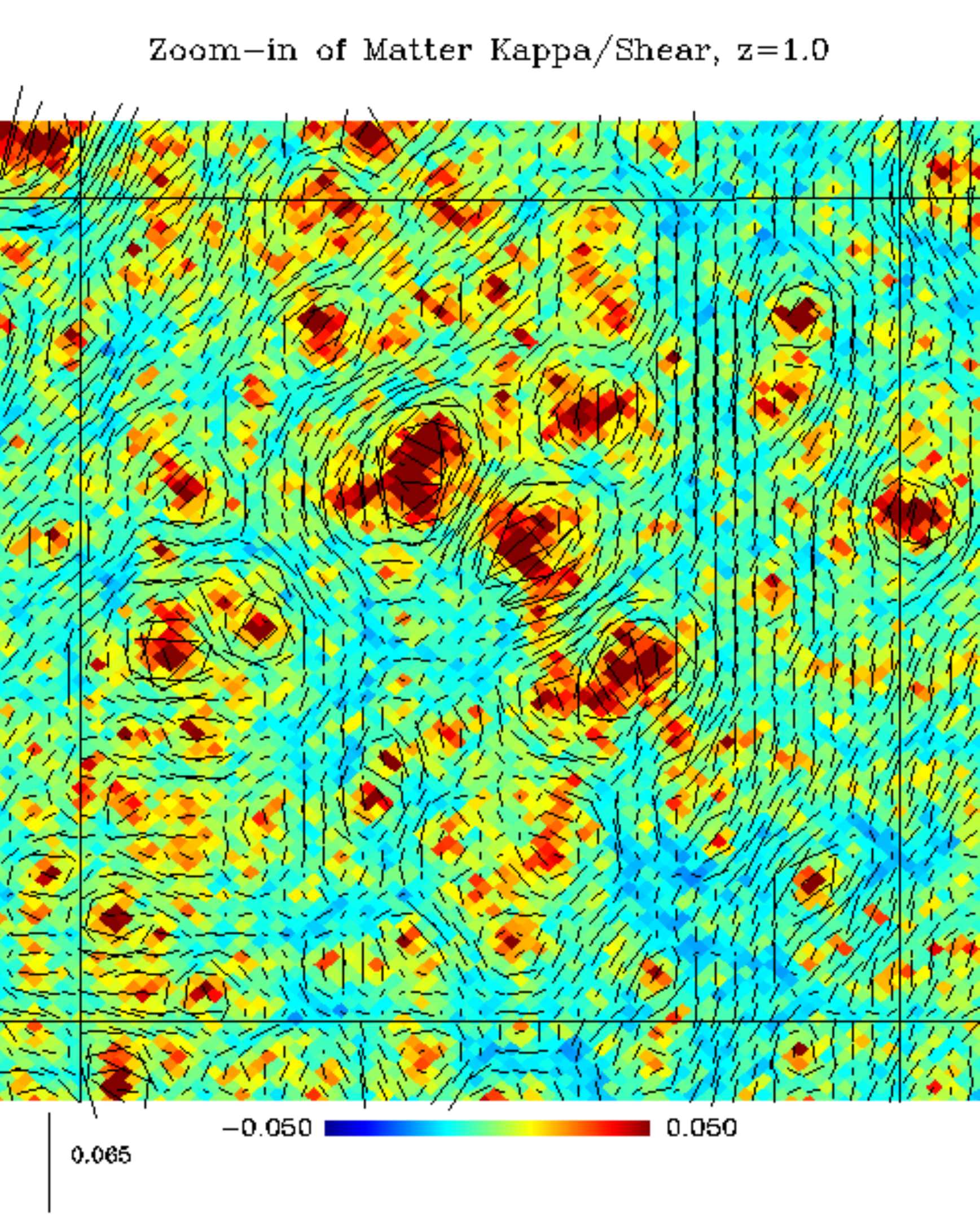}}
{\includegraphics[width=0.45\textwidth]{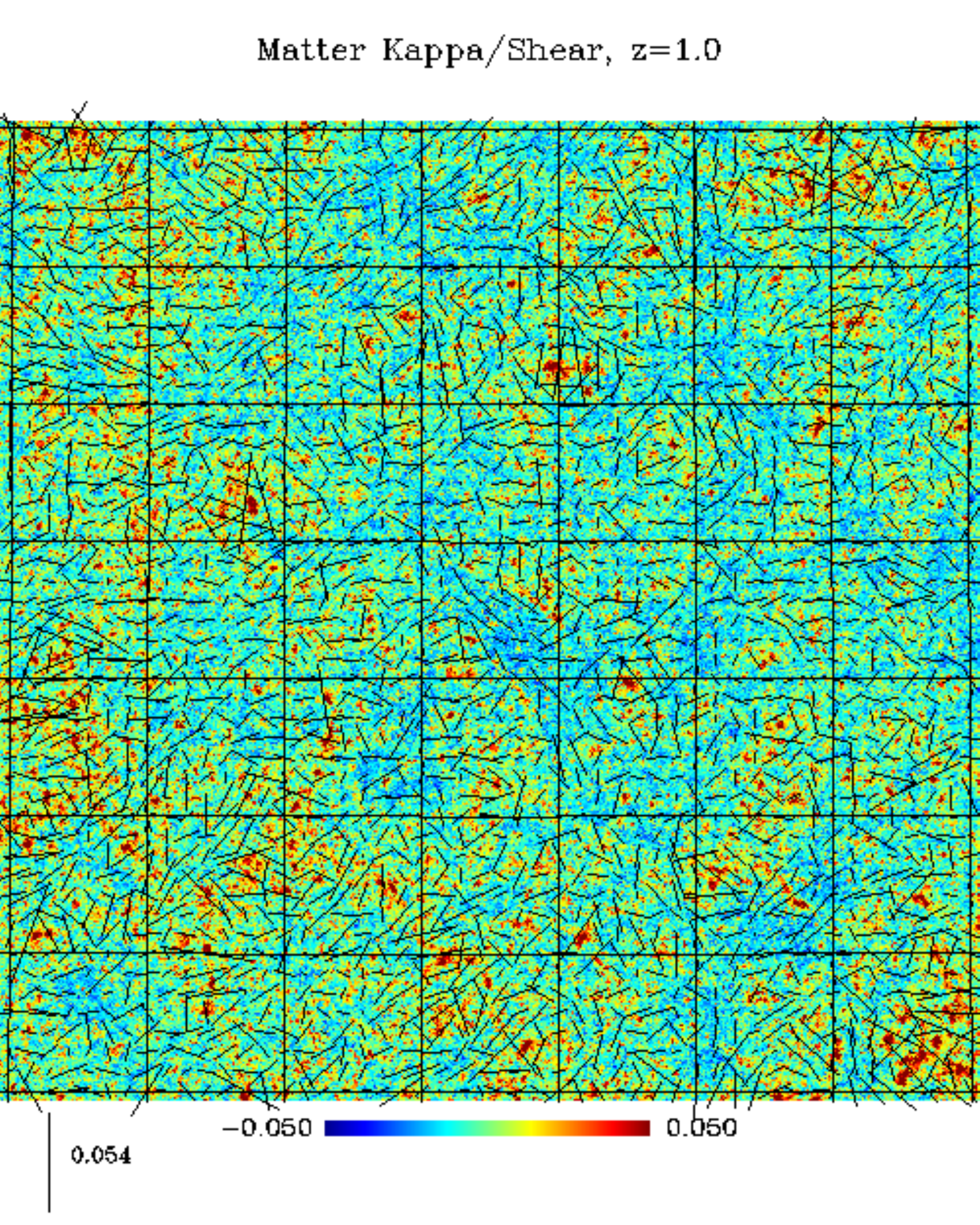}
\includegraphics[width=0.45\textwidth]{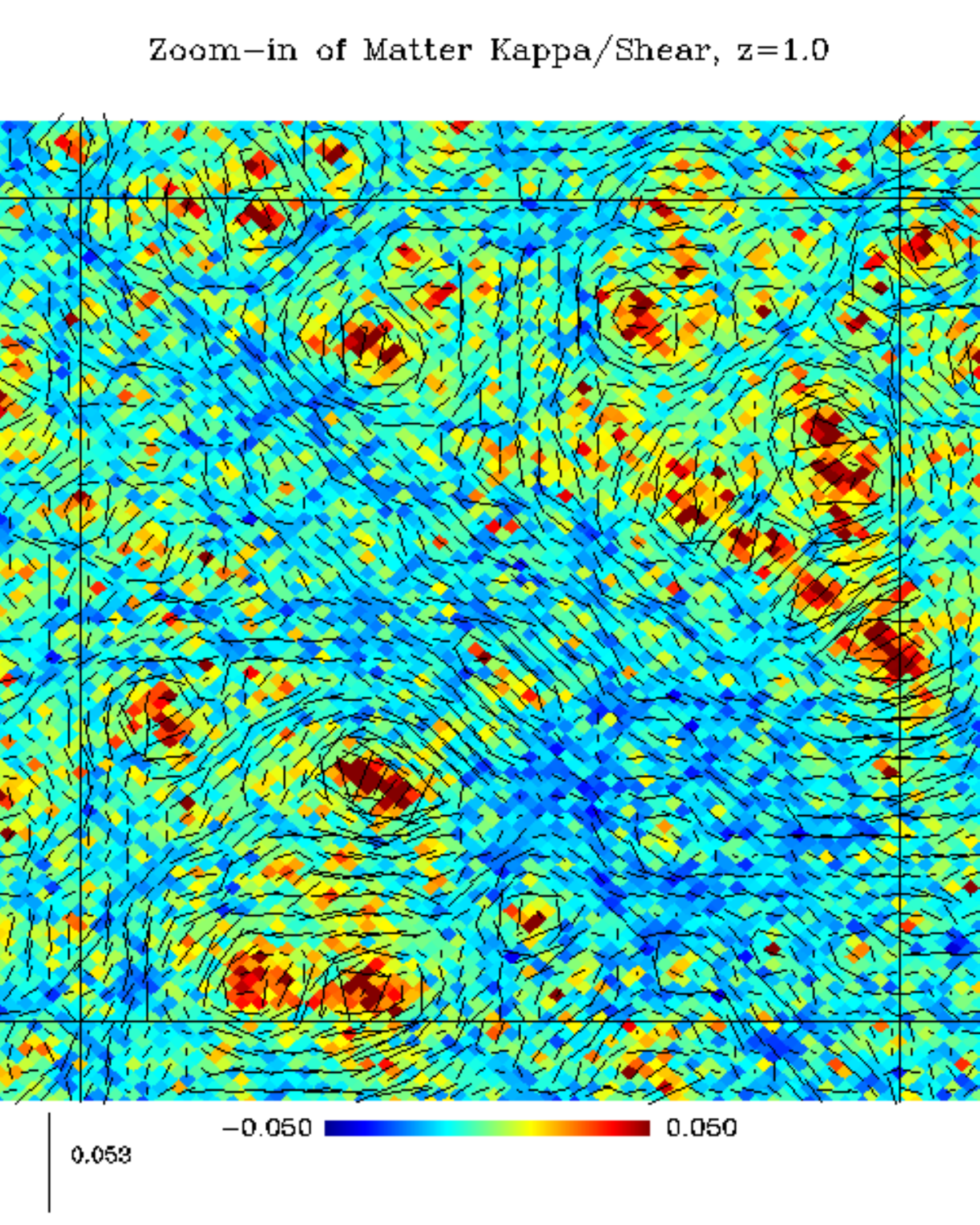}}
\caption{Top panels: (Left) 7x7 sq.deg patch showing the fluctuations in
  the convergence field (the bar at the bottom of each panel shows the
  gray scale/color code used to depict the field value, which is dimensionless) and
  shear vectors (scale for vector length displayed is shown at the bottom left; field is dimensionless)
from dark-matter ``onion shells'' of the MICE-GC simulation, for sources at
$z=1$. Rectangular grid has cells of $1$ sq.deg, corresponding to
comoving transverse lengths of $21$ Mpc/h. Shear
amplitude is given by length of the vectors, with scale as given in
the bottom left of the plot. 
(Right) Zoom-in: central 1 sq.deg grid cell of patch shown in left panel.
It shows shear vectors are tangential to matter
over-densities, as expected. 
Bottom Panels: Same as Top panels 
but for the ``Intermediate'' resolution simulation MICE-IR, which has a factor of $8$ lower mass resolution.
Mass resolution effects are reflected in the lack of small-mass halos
(or substructure in the convergence/shear maps).
This is more clearly seen in the zoom-in picture (right panels).}
\label{fig:shearmat}
\end{center}
\end{figure*}

Figure~\ref{fig:shearmat} shows 7x7 sq.deg patch of a lensing map
from the MICE-GC (top panels) and MICE-IR (bottom panels)
simulations. The rectangular grid overplot has a
side-length of $1$ deg. The amplitude of the convergence field, $\kappa$, is shown color coded,
whereas the shear field is shown with vectors on top of the
convergence amplitude. As expected in the weak-lensing regime, where the
B-mode vanishes, shear vectors are tangential to the 
projected over-densities (red spots), and point towards under-dense
(blue) regions. In the flat-sky limit, $f(\ell)\rightarrow 1$ in
Eq.(\ref{eq:shear}), so that the shear amplitude is given by the
convergence field, modulus a global sign. This theoretical expectation
is in agreement with simulation measurements, as shown in Figure~\ref{fig:clshear}, where
both power spectra converge beyond the lowest multipoles ($\ell > 10$;
see wiggly blue and black lines).  

The relative lack of resolved small-mass halos in MICE-IR (see e.g,
bottom panels in Figure~\ref{fig:shearmat} ) with respect to the higher-resolution MICE-GC (top panels) is
clearly reflected in the corresponding lower level of well-resolved
substructure in the shear maps.
This is consistent with the larger amplitude of the projected dark-matter and convergence power
spectra for MICE-GC relative to MICE-IR on the smallest scales
(\ie largest multipoles), as seen in Figure~\ref{fig:clkmassres}.
One can derive at what scales this mass resolution effect should be visible:
the smallest halos resolved in MICE-GC have a mass
$M_{min} \sim 3\times 10^{11} \Msun$ (corresponding to 10 FoF particle halos),
which is roughly the particle mass for MICE-IR.
Therefore, according to the halo model for dark-matter clustering, 
the difference in power between our high and low mass
resolution simulations should be due to the 1-halo term contribution
from halos with mass $M < M_{min}$,  not found in MICE-IR.
For the concordance LCDM cosmology adopted these halos have a size $D (M_{min})\sim$ 2 Mpc/h 
and subtend an angle of $\theta_{D} \sim 7.6$ arcmin ($\theta_{D} \sim
5.8$ arcmin) for $z=0.5$ ($z=1.0$), what projects onto multipoles $l_{D}\sim 1500$ ($l_{D}\sim 2000$).
This is consistent with what we measured in terms of mass-resolution
effects for the dark-matter clustering in Paper I (see \cite{paperI}): we found 
MICE-GC has at least $10 \%$ larger power than MICE-IR on those
scales.

Qualitatively this resolution effect is also observed in the top panels of
Figure~\ref{fig:shearmat} that shows the lensing map for
sources at $z=1$, that receives a peak contribution from lensing
halos at half the distance between the source and the observer, \ie
$z\simeq 0.5$. As argued above, at this redshift, sources with
$M<M_{min}$ which project onto an angular size $\theta_{D} < 7.6$
arcmin (i.e,  few pixel-sized structures, for the angular resolution of the maps shown),
are far more abundant than in the corresponding lensing map
for MICE-IR (see bottom panels of Figure~\ref{fig:shearmat}).

In fact, the magnitude of this effect is expected to increase with
redshift for the simple argument that low-mass halos have a relatively larger abundance at
high-redshift, as expected from hierarchical clustering, and quantified
by the evolution of the halo mass function shape with redshift (see Fig.2
in Paper II, \cite{paperII}).
Note this is also consistent with our simulation results for the
3D matter power spectra, as shown in Fig.5 of Paper I \citep{paperI}, where we
obtained a factor of $\sim 2$ larger mass resolution impact
for sources at $z=1$ with respect to those at $z=0.5$, on small
(non-linear) scales.

\section{Galaxy Lensing }
\label{sec:gallens}

Next we turn to our implementation of lensing properties of mock
galaxies using the all-sky lensing maps discussed in \S\ref{sec:lens}.
Mock galaxies are assigned using a hybrid halo occupation distribution
and abundance matching (HOD+HAM) approach, as discussed in
detail in Paper II \citep{paperII}. In the implementation used
for the galaxy mocks discussed in this series of papers, \ie
the {\tt MICECAT v1.0}, our procedure is a simple 3-step algorithm: 

\begin{enumerate}

\item{
for a given galaxy at the 3D
position in the lightcone $(\hat{\rm{n}}, z)$, where $\hat{\rm{n}}$ gives its angular
  position in the sky and $z$ its redshift, find the
  corresponding 3D pixel in the {\it discretized} lightcone, with pixel center coordinates,
  $(\hat{\rm{n}}_i, z_j)$, where the galaxy sits in (\ie the 3D pixel in the suite of
  ``onion slices'' or all-sky lensing maps in Healpix tessellation
  described in  \S\ref{sec:lens})}

\item{
get the lensing values for this 3D pixel using the
dark-matter all-sky lensing maps, $\vec{L}_{i,j}
    \equiv \vec{L} (\hat{\rm{n}}_i, z_j)$, where the components of
    the lensing vector are ${\vec L} = (\kappa, \gamma_1,\gamma_2)$ (\ie
    convergence and shear), and}

\item{
assign these pixelized dark-matter lensing values, $\vec{L}_{i,j}$, to the mock galaxy.}
\end{enumerate}

This simple implementation of galaxy lensing is limited by the pixel
resolution used, Healpix $\rm{Nside}=4096$, which corresponds to a pixel
scale of $0.85$ arcmin.  Consistently, we only expect to model lensing observables 
accurately down to $\sim 1$ arcmin scales, as we will discuss in detail 
below.   Another obvious limitation intrinsic to this
method is that different galaxies that fall within a given 3D pixel in
the ``Onion Universe''  decomposition of the lightcone, will have
identical lensing properties. These two limitations can be overcome
using the same approach but using a finer pixel scale (\ie higher
Nside) and/or using interpolation schemes for signals on the sphere
(see \eg \cite{flints}). We plan to incorporate these more accurate
implementation in future releases of the {\tt MICECAT} galaxy mocks.

\subsection{Converge autocorrelation}
\label{sec:shearcls}

In this section we shall validate the implementation of mock galaxy lensing
properties introduced in the previous section, by comparing measured
2-point correlation statistics to non-linear theory predictions.
We start by focusing on harmonic space and the angular power spectra
of lensing observables.
For this purpose, we shall use the standard Gaussian approximation to
theoretical errors, which includes sample variance and shot-noise
contributions (see \cite{kaiser92,stebbins96,crocce11}). In this
approximation we get for the variance,
\beq 
Var(C_{\ell}^{ij}) = \frac{1}{(2\ell+1) f_{sky}} \left[ (C_{\ell}^i +
n_i)(C_{\ell}^j + n_j) + C_{\ell}^{ij} \right]
\label{eq:varcl}
\eeq
 where $C_{\ell}^{ij}$ is the cross-spectra between z-bins $i$ and
 $j$, of the foreground and background lensing observables,
 respectively, and $n_i$ is
 the shot-noise contribution. For incomplete sky coverage, the
 effective reduction in independent modes available to a given
 multipole is accounted by the fraction of the sky factor, $f_{sky}$.
In the case $i=j$, we recover the variance for the auto-power spectra. 

For the convergence power spectra error-bars,  
we use the theory $C_{\ell}$'s as given by Eq.(\ref{eq:clkappa})
and the shot-noise contribution is
obtained by integrating the 3D Poisson shot-noise power of the lightcone
simulation dark-matter counts at each redshift slice,  weighted by the
 weak-lensing efficiency (\ie replacing 
$P(k,z) \rightarrow P(k,z)_{shot} = 1/\bar{n}$, being $\bar{n}$ 
the 3D dark-matter counts number density, in Eq.(\ref{eq:clkappa})
above).  The Gaussian approximation is 
expected to be accurate on large scales
and for close to all-sky surveys (see \eg \cite{cabre07}), but they
tend to underestimate errors on
small-scales, where non-linear gravitational growth induce
non-Gaussian covariances through the matter
trispectrum \citep{scoccimarro99,cooray01}. 
However projection effects inherent to lensing observables mitigate the non-Gaussian contribution relative to
the corresponding term in 3D clustering
statistics \citep{semboloni07,takada09,hilbert09}.
Given that we only have one single realization in our
analysis, the MICE-GC run, we shall stick to the Gaussian
approximation for the errors shown in this paper (unless otherwise stated)
when comparing to theory predictions, although with the obvious caveat
that they tend to underestimate covariances on small-scales. More
accurate analysis of weak lensing observables using multiple
realizations is left for future work.

\begin{figure}
\begin{center}
\includegraphics[width=0.48\textwidth]{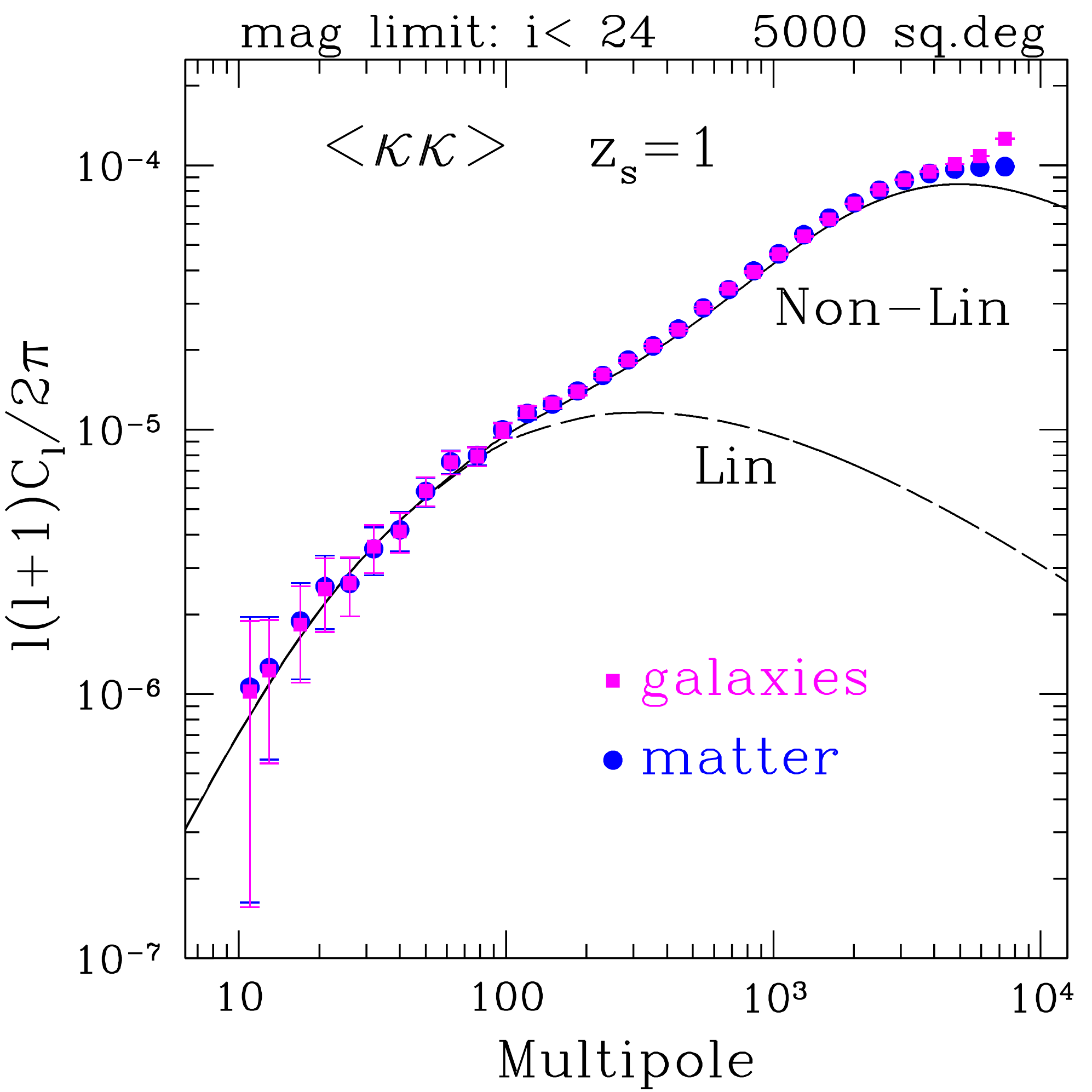}
\caption{Convergence power spectrum, $C_{\ell}^{\kappa\kappa}$, for sources at $z_s=1$, for
 matter (blue circles) and galaxies (magenta squares), compared to theory predictions.} 
\label{fig:clkk}
\end{center}
\end{figure}

Figure~\ref{fig:clkk} shows the angular power spectrum of the
convergence field measured from the MICE-GC simulation for sources at $z=1$.
The galaxy convergence field is expected to be independent of galaxy
bias, as it directly traces the underlying (projected)
dark-matter distribution. This is in fact what we observe by comparing the convergence for
a mock source galaxy sample selected with a magnitude limit in the i band, $i_{AB}<24$
(see square symbols in plot), 
and that of the underlying dark-matter density
field (circles). Theoretical errors shown are for 5000
sq.deg, which is the area of the sky of the full galaxy
mock, {\tt MICECAT v1.0}, that that we are making publicly available.

\begin{figure}
\begin{center}
\includegraphics[width=0.48\textwidth]{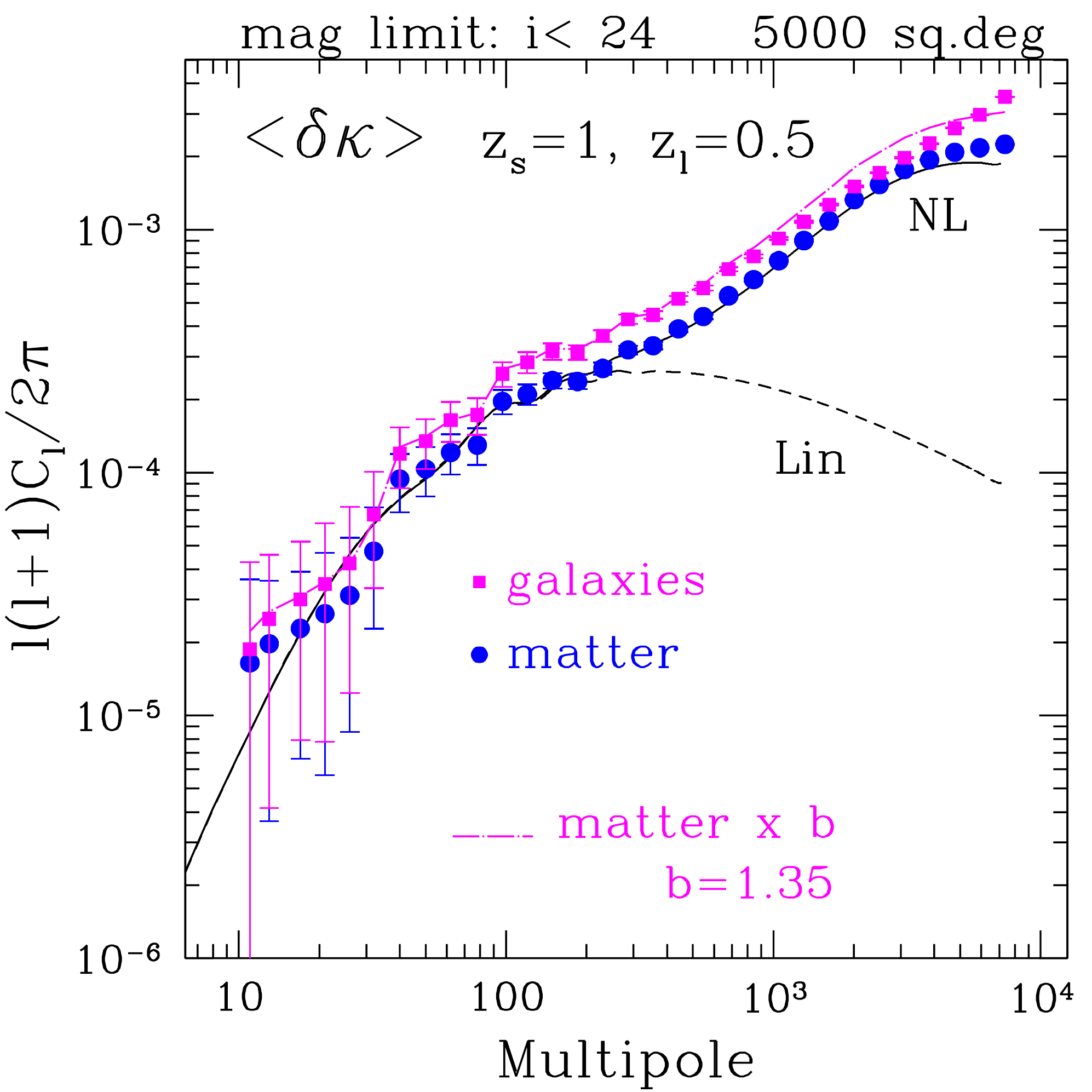}
\caption{Galaxy-Convergence power spectrum, $C_{\ell}^{\kappa g}$, for sources at $z_s=1$ and
  lenses at $z_l=0.5$. We use a redshift bin-width of $\pm 0.05$ for
  both sources and lenses. Biased matter correlation (magenta
  dot-dashed line)
  uses linear galaxy bias
  estimated from galaxy auto-correlation, Figure~\ref{fig:clgglens}.}
\label{fig:clkg}
\end{center}
\end{figure}

\subsection{Converge cross-correlation}

On the other hand, the cross-correlation between background and
foreground galaxy populations depends on the bias of the foreground
(lens) population.
Figure \ref{fig:clkg} shows that the cross-power measured for source galaxies
at $z_1=1.0$ with $i_{AB}<24$
and lenses at $z_2=0.5$ (and z-bin widths of $\pm 0.05$ for both
sources and lenses), see filled circles, is consistent with that for dark-matter
scaled by a galaxy bias factor $b\approx 1.35$ (shown by the
dot-dashed line). This is in agreement with the
bias estimate from the galaxy auto-power, as shown in the lower panel of Fig.\ref{fig:clgglens}.

\begin{figure}
\begin{center}
\includegraphics[width=0.48\textwidth]{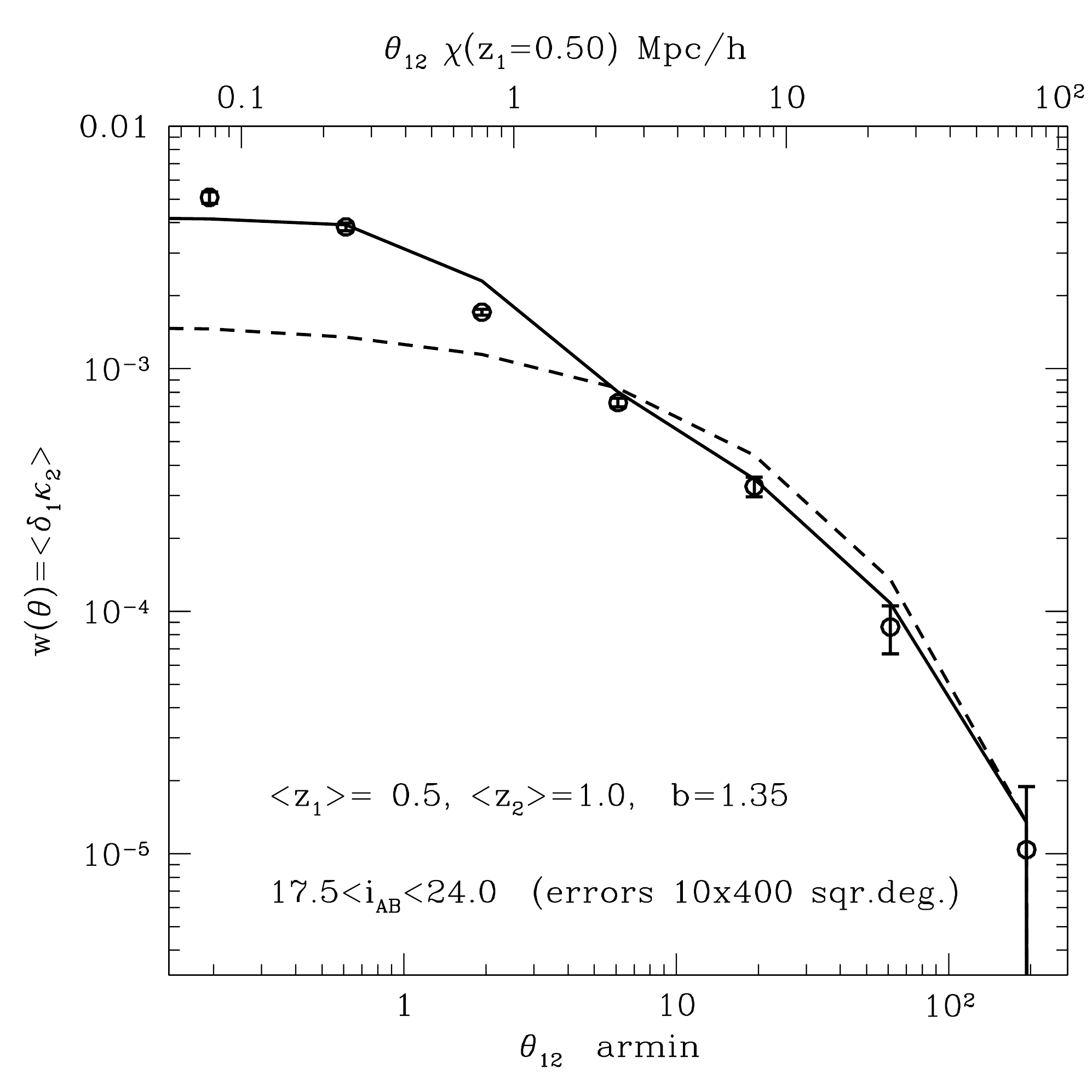}
\caption{Galaxy-Convergence cross-correlation function for sources at $z_s=1.0$ and
  lenses at $z_l=0.5$ (and z-bin widths of $\pm 0.05$ for both samples).  Dashed and continuous lines show the linear
and non-linear predictions for dark-matter. Simulation results are for $10\times400$ sq.deg. area.}
\label{fig:w2gs}
\end{center}
\end{figure}

Fig.\ref{fig:w2gs} shows the corresponding measurement in configuration space, i.e.
the  angular cross-correlation: 
\beq
w(\theta_{12})=<\delta(\theta_1,z_1)~\kappa(\theta_2,z_2)>,
\eeq
where $z_1=z_l$ correspond to the lenses and $z_2=z_s$
to the sources. We count all pairs of galaxies between the two
redshift bins and average in angular bins $\theta_{12}$
 the product of number density fluctuations (counts) and the
$\kappa$ fluctuations. We also use 
$17.5<i_{AB}<24$ so that the bias at $z_s\simeq 0.5$
 is $b \simeq 1.35$.\footnote{Note that at $z \simeq 0.5$ the MICE
   sample is only complete to $i_{AB}<22.5$, see Fig.5 in paper II.}
 The measurements agree quite well with the
 non-linear DM prediction (continuous line). On the smallest 
scales the MICE galaxies show some excess power, which can 
be interpreted as non-linear bias (similar to that also shown in 
 Fig. \ref{fig:clkg} at the largest multipoles). Note than on scales
smaller that $\theta_{12}<0.45$ arcmin we expect the MICE
measurements to become flat because this is within the
healpix pixel radius in  the $\kappa$ maps. This corresponds
to $\simeq 100$ Kpc/h for $z_l \simeq 0.5$ (see top label
in the figure), where the non-linear DM prediction also 
flattens down and the MICE simulations approaches the
softening lengths of $50$ Kpc/h. We conclude from this
that the resolution of the $\kappa$ maps is adequate to
model weak lensing to sub arcminute scales. Errors in
Fig~\ref{fig:w2gs} are obtained from 10 patches of 400 sq.deg.

\begin{figure}
\begin{center}
\includegraphics[width=0.48\textwidth]{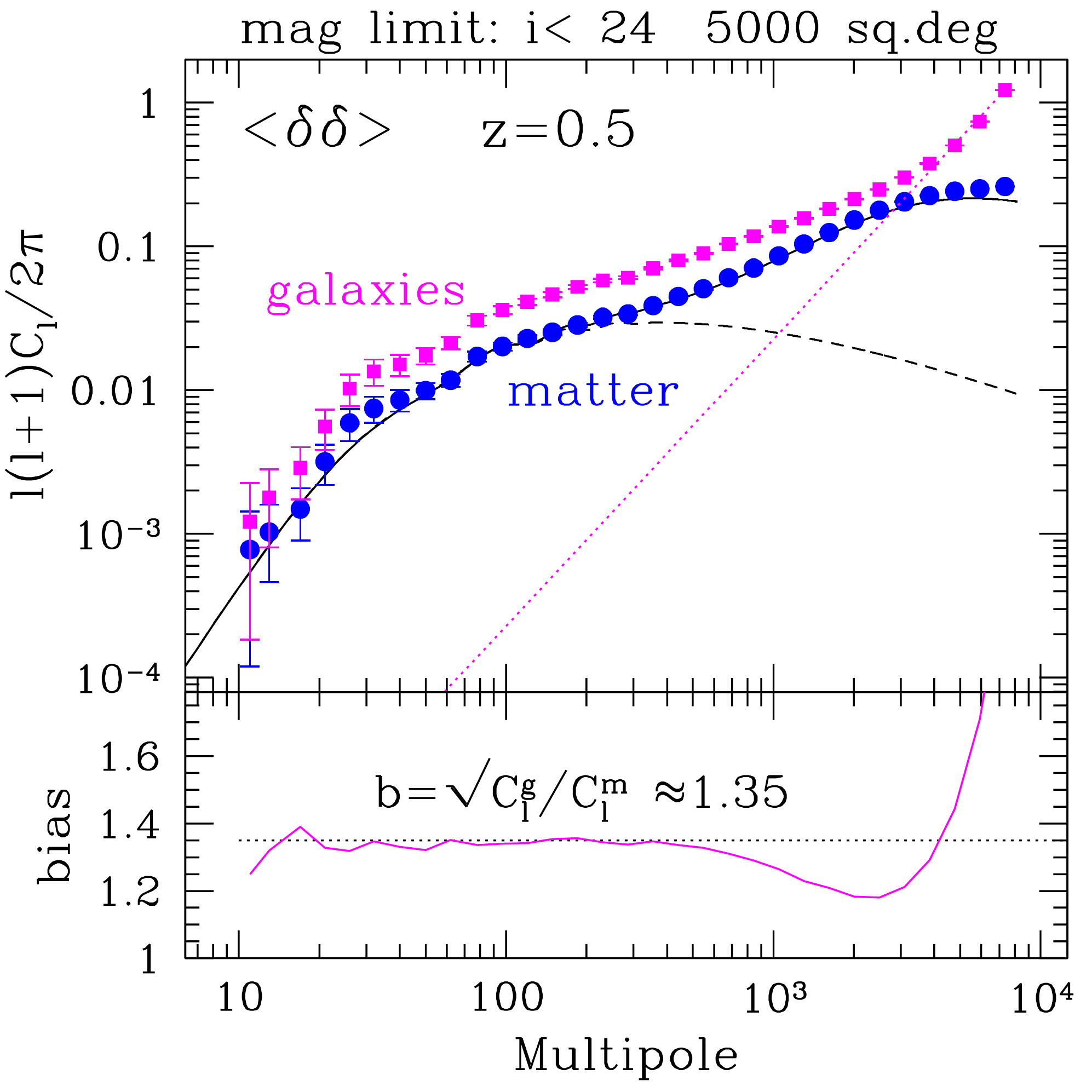}
\caption{Galaxy auto-power spectra for matter (blue circles) and
  galaxies (magenta squares) at $z=0.5$, compared to linear (dashed) and
  non-linear theory (solid line). We use a z-bin width of $\pm 0.05$. Lower panel shows the galaxy bias, given by the square root
  of the ratio between the galaxy and matter power spectra. A scale-independent
 (linear bias) fit to this ratio, $b\simeq 1.35$, is also shown for reference (dotted line). 
}
\label{fig:clgglens}
\end{center}
\end{figure}

\subsection{Shear correlation functions}
\label{sec:shear2pcf}

Next we validate the shear implemented in MICE dark-matter outputs and
galaxy mocks by computing shear correlation functions in
configuration space. For comparison with theory predictions, we shall
use the Legendre transform of the Gaussian error-bars used for the
angular power spectra, Eq(\ref{eq:varcl}) (see also Eq.(18) in \cite{crocce11}), what is a good approximation
for large enough scales and nearly all-sky surveys (\ie provided individual multipoles in the
angular power spectra are uncorrelated). We emphasize that Gaussian
errors are only used to give an idea of the size of the uncertainties
involved in the lensing observables discussed, rather than providing
an accurate error estimate.

\begin{figure}
\begin{center}
\includegraphics[width=0.48\textwidth]{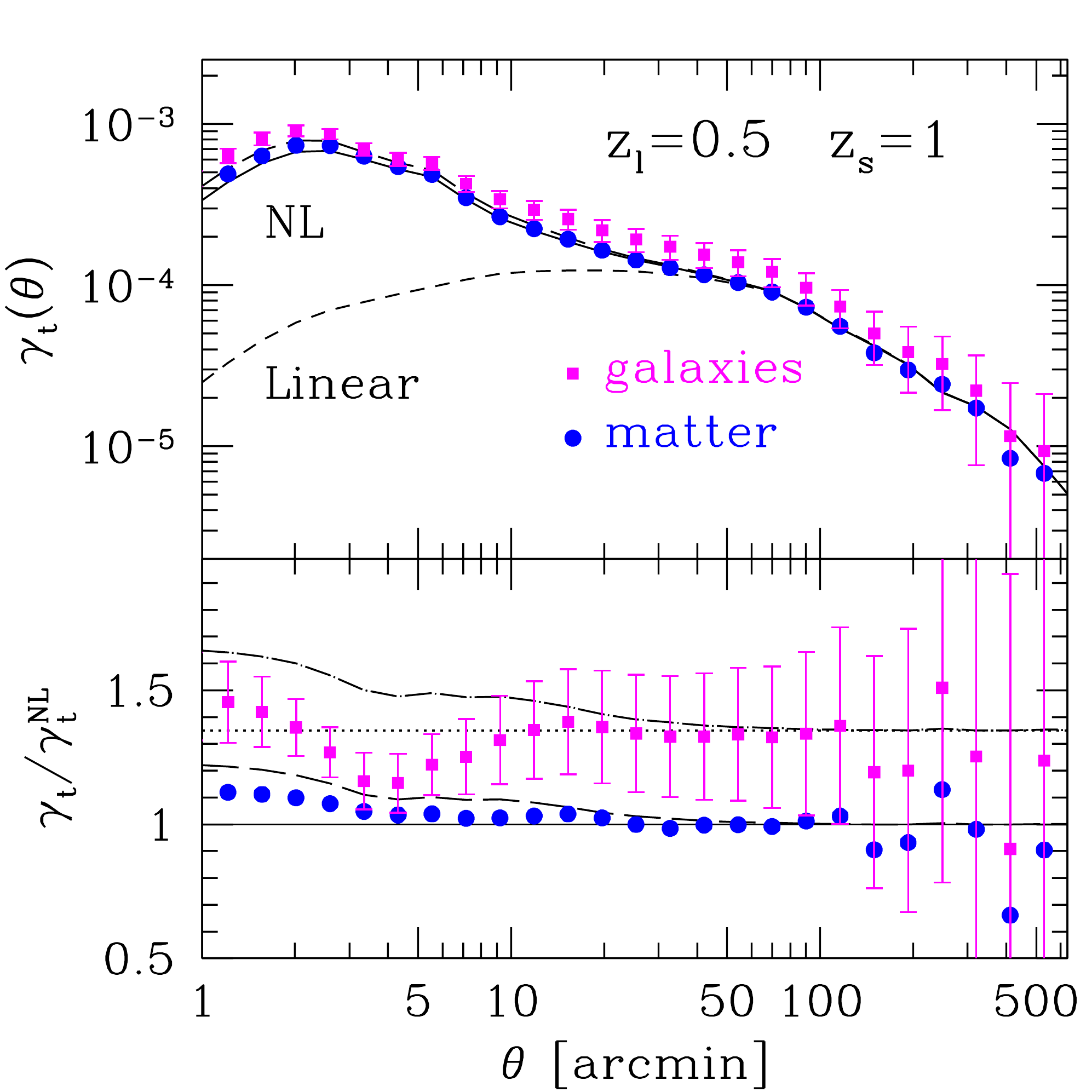}
\caption{Tangential shear for sources at $z=1$ and lenses at
  $z=0.5$. Lower panel shows deviations with respect to non-linear
  theory (Halofit) predictions.}
\label{fig:gammat}
\end{center}
\end{figure}

The average tangential shear of a background galaxy population, $\gamma_t$,  is
directly related to the cross power spectrum of the convergence field
of the background galaxies and the foreground galaxy number
counts, $C_{\ell}^{\kappa g}$ (see \cite{jeong09}),
\beq
\gamma_t(\theta) = \frac{1}{2\pi} \int d \ell \,
\ell \,
J_{2}(\ell \theta) \, C_{\ell}^{\kappa g}
\label{eq:gammat}
\eeq 
where we have taken the flat-sky limit, which is very accurate for
practically all angular scales (\ie $\theta <$ few degrees). The exact
expression can be obtained by simply replacing the Bessel functions
above by Legendre polynomials (see \cite{deputter10}).
In Fig.\ref{fig:gammat} we show results for the tangential shear
measured in the dark-matter and the mock galaxies, 
for a source galaxy sample at $z_s=1$ and a lens
population at $z_l=0.5$. Non-linear effects, that become important at
$\theta < 100$ armcin, are well captured in the simulations.
In the lower panel of Fig.\ref{fig:gammat} we see that MICE-GC
dark-matter has
more power than predicted by Halofit on non-linear scales, but it
shows a lack of power relative to the revised Halofit prediction, what
is consistent with our findings for the convergence power spectrum,
Fig.\ref{fig:clkmassres}. We also show the result of scaling up both the
Halofit (dotted) and revised Halofit (dot-dashed) predictions, with the linear galaxy bias,
$b\simeq 1.35$, estimated form the lens galaxy population (see
Fig.\ref{fig:clgglens}). Our measurement of the galaxy tangential
shear is in rough agreement with the {\it linearly biased} Halofit
prediction, although they disagree on the smallest scales, where the simple
linear bias assumption seems to break down.

On the other hand, a common observable in lensing surveys is the 2-point shear
correlation functions, $\xi_{\pm}$, which in turn are related to the
tangential and cross-component of the shear,
\beq
\xi_{\pm} (\theta) = <\gamma_t\gamma_t> \pm
<\gamma_{\times}\gamma_{\times}>(\theta)
\eeq
 where, $\gamma_t = {\cal R}e (\gamma\e^{-2i\theta})$, and
 $\gamma_{\times} = {\cal I}m (\gamma\e^{-2i\theta})$,
and $\theta$ is the polar angle of the separation vector,
${\pmb \theta}$.

\begin{figure}
\begin{center}
\includegraphics[width=0.48\textwidth]{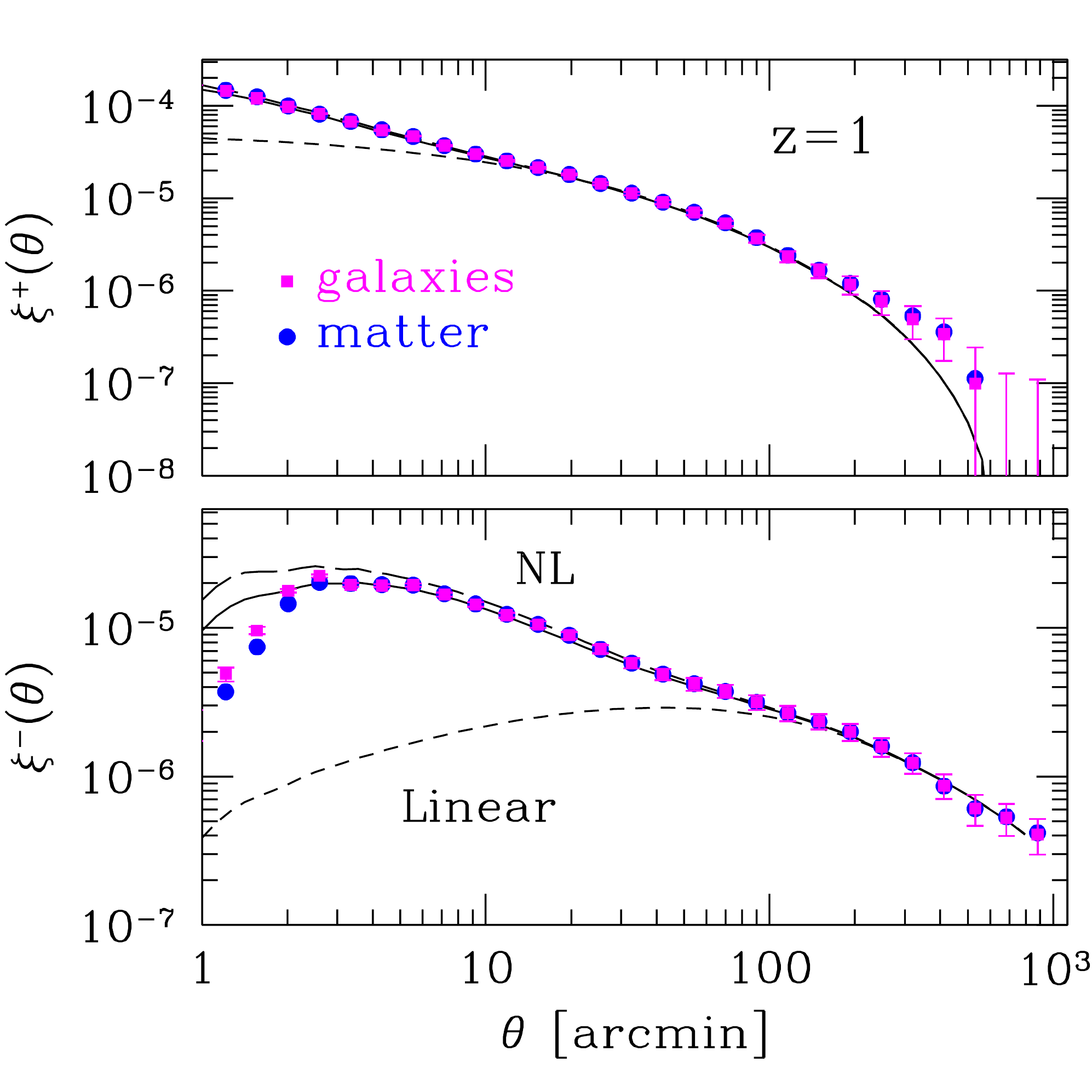}
\caption{Shear correlation functions, $\xi^{+}$ (top) and $\xi^{-}$
  (bottom panel)
  for sources at $z=1$. Linear and non-linear theory predictions are
  shown with dashed and solid lines, respectively (see text for details).}
\label{fig:shear2pcf}
\end{center}
\end{figure}

In the weak lensing limit (\ie in the
absence of rotational modes) these shear correlations are related to
the the gradient or E-mode component of
the shear tensor (see \cite{bartelmann01}),
\beq
\xi_{\pm}(\theta) = \frac{1}{2\pi} \int d \ell \,
\ell \,
J_{0/4}(\ell \theta) \, C_{\ell}^{\gamma \gamma}
\label{eq:shear2pcf}
\eeq 
where we have assumed the Limber approximation, for which its also
true that the shear and convergence power spectra are identical,
$C_{\ell}^{\gamma\gamma} =  C_{\ell}^{\kappa\kappa}$,
as explicitly shown in Fig.\ref{fig:clshear}. We can thus compute
$\xi_{\pm}$ by using the convergence power
spectra. Note that this in turn is saying that both shear correlation
functions are not independent, as they are both related to the same underlying 
lensing potential. In fact, one can easily invert
Eq.(\ref{eq:shear2pcf}) to get (\eg \cite{schneider03}),
\bea
C_{\ell}^{\kappa\kappa} &=& 2\pi \int^{\infty}_{0} d\theta
\,\theta\,\xi_{+}(\theta) J_{0}(\ell \theta) \nonumber \\
&=& 2\pi \int^{\infty}_{0} d\theta\,\theta\,\xi_{-}(\theta) J_{4}(\ell \theta) 
\label{eq:clfromxi}
\eea
Figure~\ref{fig:shear2pcf} shows the shear 2-point
correlation functions for sources at $z=1$. For $\xi_{+}$ we see that
both MICE-GC mock galaxies and dark-matter are in close agreement with 
the non-linear prediction, as they deviate from
the purely linear regime for $\theta < 10$ arcmin, and down to $\approx
1$ arcmin scales. 
In particular, predictions from Halofit (solid line) and the {\it revised} Halofit
(long-dashed) can
be hardly distinguished on these non-linear scales.
The agreement between the simulation and theory 
is even more remarkable for $\xi_{-}$, for which non-linear effects
become significant at much larger scales, $\theta < 100$
arcmin,  in much the same way it happened for the tangential shear, 
Fig.\ref{fig:gammat}. 

\section{Magnification from Galaxy cross-correlations}
\label{sec:magbias}

Gravitational lensing by large-scale structures in the universe
changes the number density of background sources and thus
it induces a cross-correlation signal between background and
foreground galaxy populations \citep{moessner98,bartelmann01}. Such
cross-correlations have been measured using samples of distant quasars
magnified by low redshift galaxies (\eg \cite{benitez97,gaztanaga03,myers05,scranton05}), 
that can be used to put constraints on the galaxy-mass power
spectrum \citep{jain03}.

\begin{figure}
\begin{center}
\includegraphics[width=0.48\textwidth]{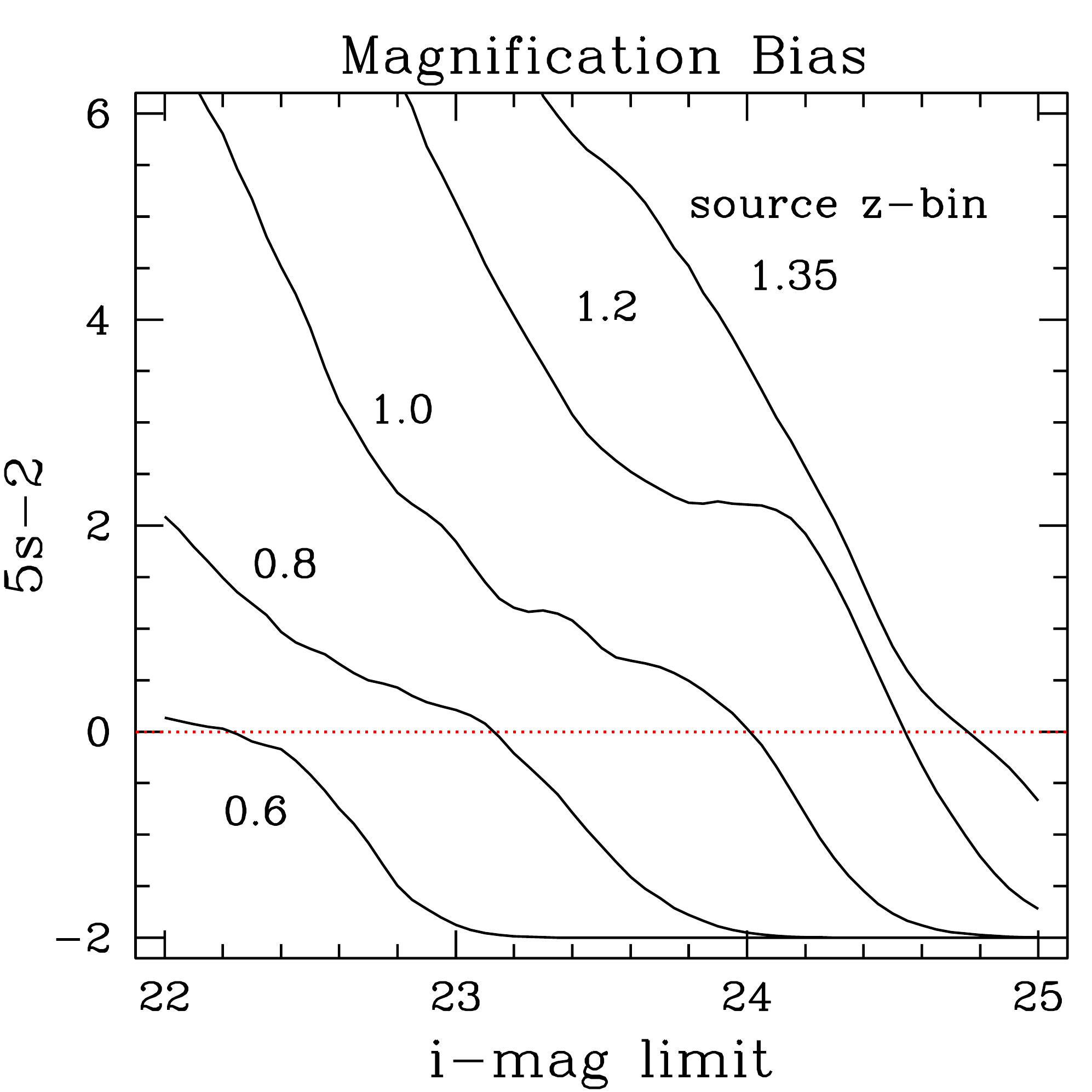}
\caption{Magnification bias for magnitude limited source galaxy samples of the
  MICE-GC mocks. Samples have a z-bin width of $\pm 0.05$.
Note that at $z \simeq 0.5$ the MICE 
 sample is only complete to $i_{AB}<22.5$, see Fig.5 in paper II. This
 is why $s$ tends to zero for low redshifts and faint magnitudes.}
\label{fig:magbias}
\end{center}
\end{figure}

For a magnitude limited survey,  the cumulative number of galaxies
above a flux limit $f$ scales as $N_{0}(>f) \sim A f^{\alpha}$, where $A$
is the area of the survey, and $\alpha $ is the power-law slope of
the background number counts. Lensing preserves the surface
brightness of galaxies by increasing the observed survey depth
(i.e, decreasing the effective flux limit) and the effective
survey area by the same amount:  $f \rightarrow f/\mu$, $A \rightarrow A/\mu$,
where $\mu$ is the magnification.
These two competing effects induce the so-called {\it magnification
  bias} in the cumulative number of background sources,
\beq
N(>f) \sim  \frac{1}{\mu} A \left(\frac{f}{\mu}\right)^{-\alpha} = \mu^{\alpha-1} N_{0}(>f)
\label{eq:cumnumgal}
\eeq
In the weak-lensing limit, $\mu = 1 + \delta_{\mu}$ where
$|\delta_{\mu}| \ll 1$, and we can Taylor expand, $\mu^{\alpha-1} \approx 1 + (\alpha
-1) \delta_{\mu}$ and therefore the magnified overdensity of
background sources is given by,
\bea
\delta_{all} &=& \frac{N-N_{0}}{N_{0}} = \delta_m + \delta_p \nonumber
\\
&=&  (\alpha-1)\delta_{\mu} = (2.5s-1)\delta_{\mu} = (5s-2)\delta_{\kappa}
\label{eq:deltanumgal}
\eea
where in the last equality, we have used the simple relation between
the fluctuations in magnification and convergence of dark-matter counts,
$\delta_{\mu} = 2\,\delta_{\kappa}$, that is valid in the weak-lensing limit.
Note that in $\delta_{all}$  we have defined the two qualitatively
different contributions:
\begin{enumerate}
\item{{\it counts from magnified magnitudes}, 
\beq
\delta_m = \alpha \,\delta_{\mu}
\label{eq:magmag}
\eeq}
\item{{\it counts from  lensed positions}
\beq
\delta_p =-\delta_{\mu}
\label{eq:magpos}
\eeq}
\end{enumerate} 
These two contributions cannot be separated observationally, but we
define two different galaxy samples accordingly in our simulation in order to validate
the two magnification contributions separately.
Above, we have defined the the logarithmic slope of the background
number counts at redshift $z$, for a magnitude limit $m$,
\beq
s = 2\alpha/5 \equiv \frac{ d {\rm Log_{10}}N(<m,z)}{dm}.
\eeq 
The net magnification from these two competing effects depends on
how the loss of sources due to the area dilution, $\delta_p$, is compensated by the
gain of sources from the flux magnification, $\delta_m$. Number counts
for source populations with flat luminosity functions, such as faint
galaxies, decrease due to magnification, 
whereas sources with steep luminosity functions, such as
quasars, increase.
Note that, in the particular case when $s=0.4$, then $\alpha=1$, and there is no net
magnification effect.

\subsection{Implementation in galaxy mocks}

Below we describe how to implement magnification in the magnitudes and
positions of mock galaxies. 

\begin{enumerate}
\item{{\it magnified magnitudes:}
flux magnification makes the
mock galaxy magnitudes, $m$, brighter by an amount,
\beq
\Delta m = \frac{5}{2} {\rm Log_{10}}\mu = 2.5\, {\rm
  Log_{10}}(1+\delta\mu) \simeq \frac{5}{\ln{10}} ~\kappa
\label{eq:deltam}
\eeq
where in the last equality we have Taylor expanded $ {\rm
  Log_{10}}(1+\delta\mu)$ and used the fact that $\delta\mu \simeq 2\kappa$ in the weak-lensing limit.
Therefore, knowing the value of the convergence, $\kappa$, at a given
point in the source plane, it is straightforward to compute the flux
magnification induced, that in turn produces the change in the
background number counts, $\delta_m$.} 

\item{{\it magnified or lensed positions:} the ``observed'' or lensed
position, ${\pmb\beta}$, of a light ray is shifted from the ``true'' or unlensed
position, ${\pmb\theta}$, by an angle given by the deflection vector,
${\pmb\alpha}$, according to the {\it lens equation} on the source
plane (see \eg \cite{bartelmann01}).
In the single-plane (or Born) approximation, 
the lens equation reads,
\beq
{\pmb\theta} = {\pmb\beta} + {\pmb\alpha}
\label{eq:lenseq}
\eeq 
where the deflection vector, ${\pmb\alpha}$ is a tangent vector at the
{\it unlensed position} of the light ray, and the lensed position is
found by moving along a geodesic on the sphere in the
direction of this tangent vector and for an arc length given by the deflection angle,
$\alpha$. If we denote the unlensed position on the sphere by
$(\theta,\phi)$, then the lensed position, $(\theta^{\prime},\phi +
\Delta \phi)$, can be simply derived by using identities of spherical
triangles \citep{lewis05,das08},
\bea
\cos\theta^{\prime} &=&
\cos\alpha\,\cos\theta-\sin\alpha\,\sin\theta\,\cos\delta \nonumber \\
\sin\Delta\phi &=& \sin\alpha\,\sin\delta/\sin\theta
\label{eq:remap}
\eea 
where the (complex) deflection vector is projected on the polar basis of the
sphere, $(\vec{e_{\theta}},\vec{e_{\phi}})$,  at the unlensed position as, 
\bea
\vec{\alpha} &=& \alpha_{\theta}\,\vec{e_{\theta}} +
\alpha_{\phi}\,\vec{e_{\phi}}  \nonumber \\
&=& \alpha\,\cos\delta \,\vec{e_{\theta}} + \alpha\,\sin\delta\,
\vec{e_{\phi}} = {\cal R}e (\alpha) \,\vec{e_{\theta}} + {\cal I}m (\alpha)\,\vec{e_{\phi}}
\eea
being $\delta$ the angle between the deflection vector and the polar
basis vector $\vec{e_{\theta}}$.
We use Eq.(\ref{eq:remap}) above to {\it re-map} source galaxy positions
due to the lensing by the large-scale structure in the lightcone Nbody
simulation.}

\end{enumerate}

\subsection{Validation}

We consider two different observable sources for magnification: counts
and magnitudes. In both cases we correlate foreground density (counts)
fluctuations. In the former we also use counts for the background
while in the later we use fluctuations in the background magnitudes.

\subsubsection{Counts}

\begin{figure*}
\begin{center}
{\includegraphics[width=0.45\textwidth]{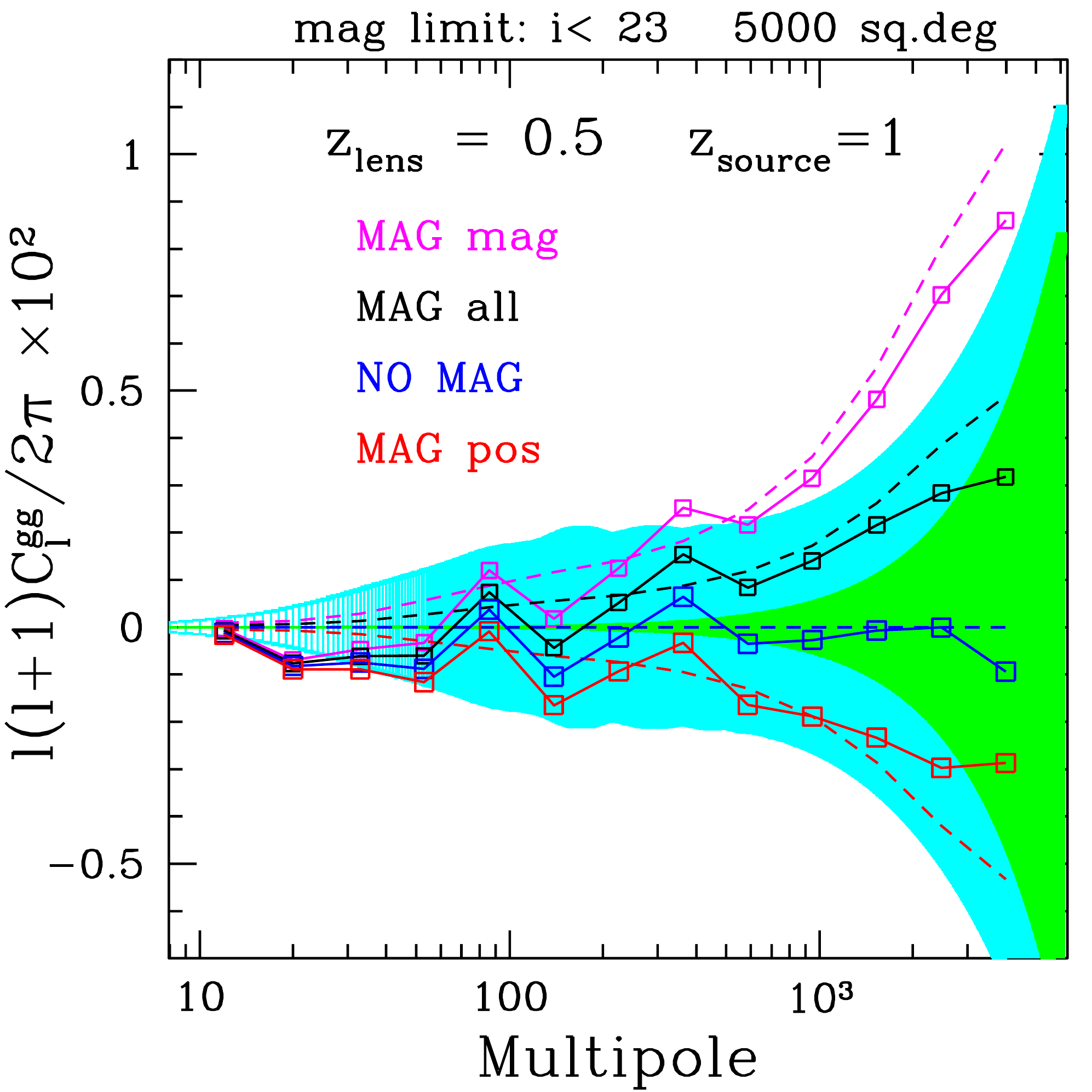}
\includegraphics[width=0.45\textwidth]{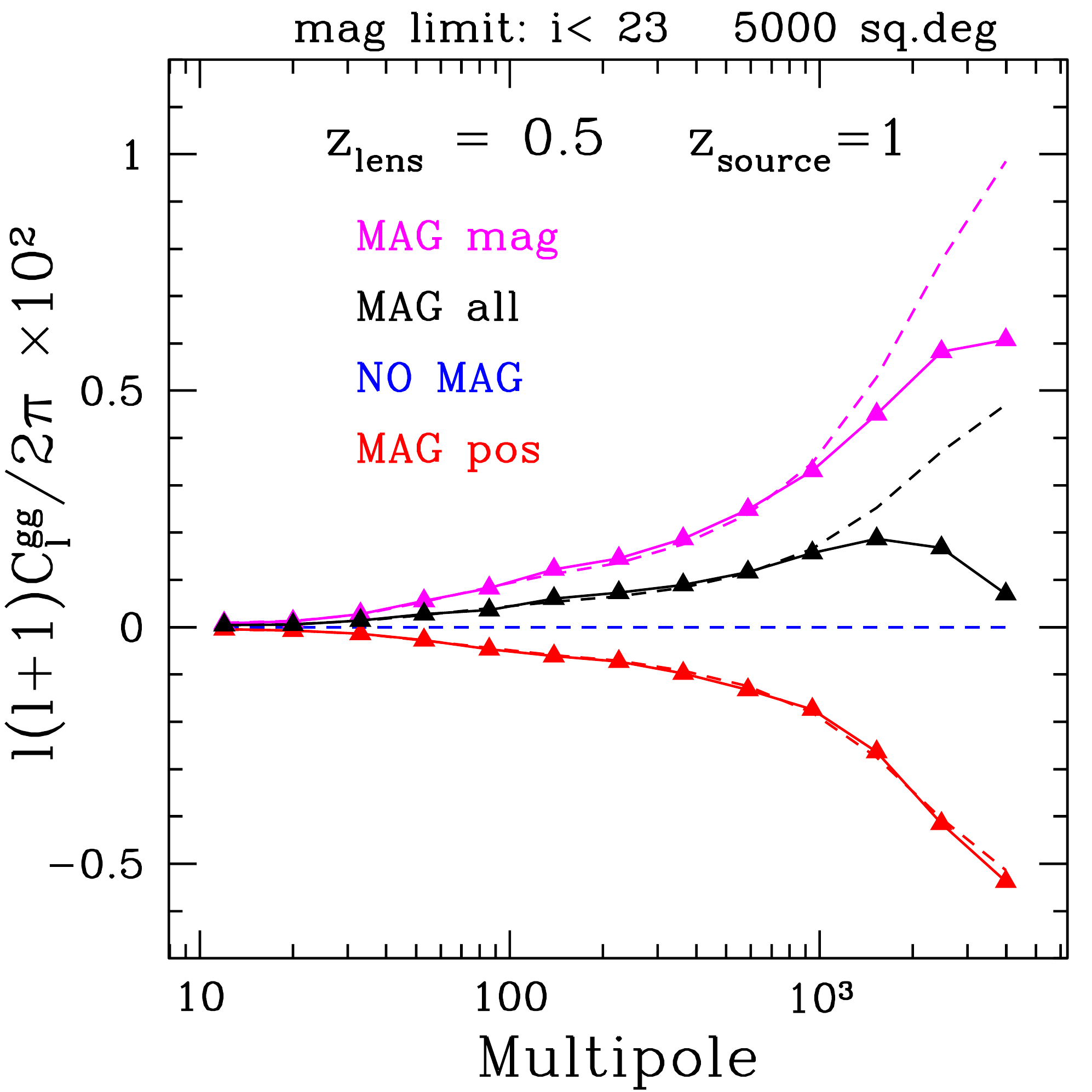}}
{\includegraphics[width=0.45\textwidth]{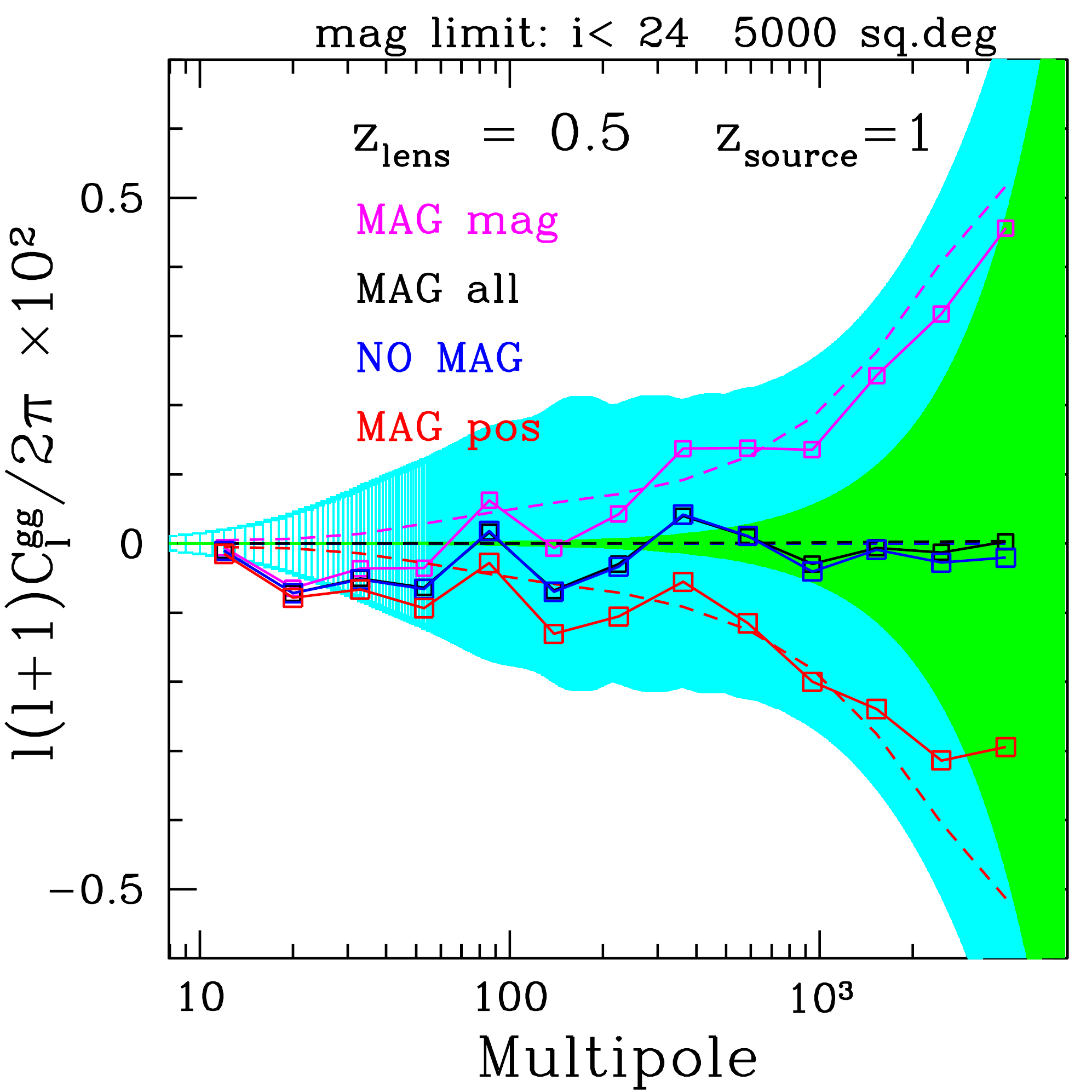}
\includegraphics[width=0.45\textwidth]{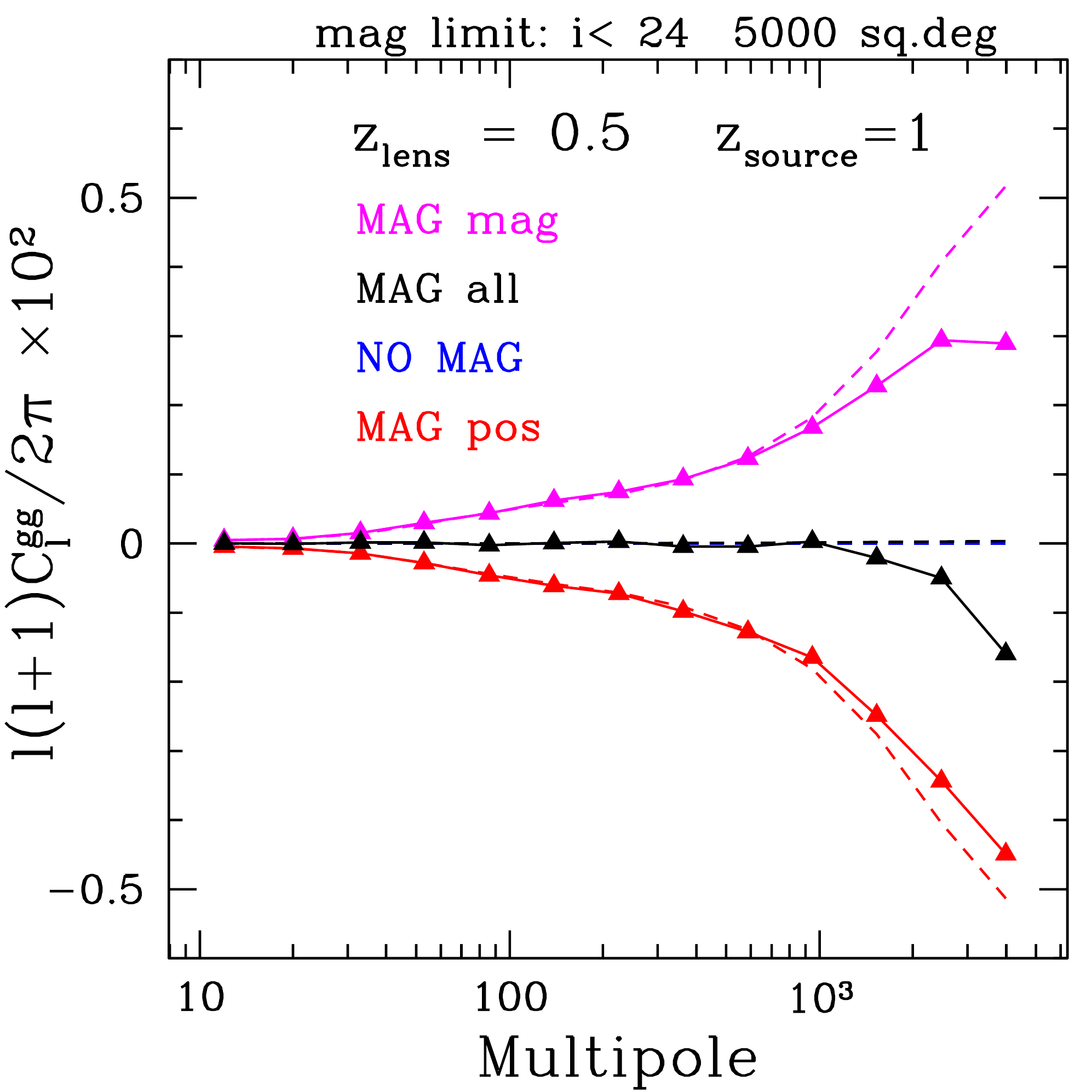}}
\caption{Magnification bias from cross-correlations of background sources at $z_s=1$ and
  lenses at $z_l=0.5$, for magnitude limited samples. Top panels show
  the case for $i<23$ whereas bottom panels illustrate the ``null
  magnification test'', $i_{AB}<24$ for this cross-correlation pair. 
Four magnitude limited mocks are constructed with
  and without magnification in the magnitudes and/or positions of
  galaxies. {\bf Left panels}: Dashed lines show theoretical predictions, whereas open symbols
  show simulation measurements. A single cross-correlation pair is noise
  dominated: 1-$\sigma$ 
errors are shown by shaded areas (shot-noise in green, shot-noise
plus sample variance in cyan shaded).  {\bf Right panels}: filled symbols show differences in
  power spectra that largely cancel out sample variance that
  dominates all cross-correlation measurements (see text for details).}
\label{fig:clkg_mag}
\end{center}
\end{figure*}

In order to validate the magnification signal in the simulation, we
have selected a magnitude limited sample from the parent HOD mock, by
imposing a cut in the $i _{AB}$ band (which is in principle less affected by
non-cosmological signals such as dust extinction), and
have constructed four different cross-correlation pairs between
foreground counts (lenses) and background counts (source) galaxies, well separated
in redshift:
\begin{flalign*}
&& <\delta(z_l)~\delta_{m}(z_s)> && \text{magnified magnitudes}, &&\\
&& <\delta(z_l)~\delta_{p}(z_s)> &&\text{magnified positions},   &&\\
&& <\delta(z_l)~\delta_{all}(z_s)> &&\text{magnified magnitudes and positions}, &&\\
&& <\delta(z_l)~\delta_{nomag}(z_s)> &&\text{no magnification},
\label{eq:crosspairs}
\end{flalign*}
where $\delta_{m}$ and
$\delta_p$ are given by Eqs.(\ref{eq:magmag})-(\ref{eq:magpos}) respectively.
For both the predictions and simulations we take:
$\delta(z_l) \approx \delta_{nomag}(z_l)$, 
so that lenses are at a sufficiently low redshift 
that they are negligibly magnified. 
In fact, lens magnification
contributes to the cross-correlation signal at the percent level only for
at $z_l<0.5$ \citep{ziour08}.
Therefore the cross-correlation
between (magnified) source and (un-magnified) lens galaxy populations are given by
(see Eq.(\ref{eq:deltanumgal}) ,
\bea
<\delta(z_l)~\delta(z_s)> &=& <\delta(z_l)~\delta_{all}(z_s)>  \nonumber \\
&=& b\,(5s-2)<\delta_m(z_l)~\delta_{\kappa}(z_s)>
\label{eq:ggcross}
\eea
We can thus predict the magnification signal expected by using the
non-linear matter cross-correlations,
$<\delta_m(z_l)~\delta_{\kappa}(z_s)>$ (we use Halofit as implemented
in {\tt CAMB sources}), scaled by the factor, $b\,(5s-2)$,
where $b$ is the linear bias
derived from the auto-correlation of the lens galaxy
population (see Fig.~\ref{fig:clgglens}), and  $(5s-2) $ the magnification bias
of the source galaxy sample (see Fig.~\ref{fig:magbias}).

Figure~\ref{fig:clkg_mag} shows the measurements in MICE-GC for these four
pairs separately, for lenses located at $z_l=0.5$, ie., the peak of the weak-lensing efficiency
for sources at $z_s=1$. We choose bin widths $\Delta z=0.1$, centered at
$z_l$ and $z_s$, to include
enough galaxies in each source/lens bin to minimize the impact of
shot-noise in the cross-correlation errors. 
We display two illustrative cases: top panels of
Fig.\ref{fig:clkg_mag} show the case for $i_{AB}<23$ for which theory
predicts a magnification bias $(5s-2) \approx 2$ (see Fig.\ref{fig:magbias}), and thus a large positive
 cross-correlation induced by magnification. This is further enhanced
 by the galaxy bias factor of the lenses, as the galaxy
 cross-correlations are given by on Lower panels show a {\it null
 magnification test}, \ie the case when the magnification bias vanishes.
Overall we find good agreement between theoretical
predictions (dashed lines) and simulations (open
symbols) in both test cases. Magnification from a single cross-correlation pair is noise
dominated for all the dynamical range: 1-$\sigma$ statistical Gaussian
errors are shown by shaded areas. Shot-noise (inner green shaded)
dominates on small-scales (\ie high multipoles) whereas sample
variance dominates the total error (cyan shaded) on large scales
(\ie low multipoles).

In order to suppress the sample variance
contribution that dominates the cross-correlations on basically
all-scales, we shall take
the difference in cross-correlation pairs that {\it see} approximately
the same sky realization. 
For example, to estimate the {\it
  magnified positions} (MAG mag) pair above without sample-variance,
we equate,
\beq
 <\delta(z_l)\delta_{mag}(z_s)> = <\delta(z_l)\delta_{all}(z_s)>
 -<\delta(z_l)\delta_{pos}(z_s)>
\label{eq:svfreepair}
\eeq 
This is shown to bring simulation measurements
in much better agreement with theory predictions for all the
dynamical range. The {\it  cosmic-variance-free measurement},
Eq.~(\ref{eq:svfreepair}), is shown in the right panels of Figure~\ref{fig:clkg_mag},
as compared direct cross-correlation pair (top panel), down to the scales that are
 affected by the pixel scale of the maps, $\ell\sim 10^4$.
Other cosmic-variance-free pairs are also shown in the same figure
(see filled symbols). Note however that this trick only works
approximately: one small-enough scales (\ie few arcmins), the lensed
sky is different than the unlensed one (\ie due precisely to the
lensing of background sources) and thus the sample variance affecting
lensed and unlensed galaxy populations is slightly different. This
could explain why on few arcmin scales, the cosmic-variance free
measurement to be in disagreement with theory expectations for each of
the samples cross-correlated. A more detailed analysis of simulated
magnification measurements from cross-correlations of galaxies will be
presented elsewhere.

\begin{figure}
\begin{center}
\includegraphics[width=0.48\textwidth]{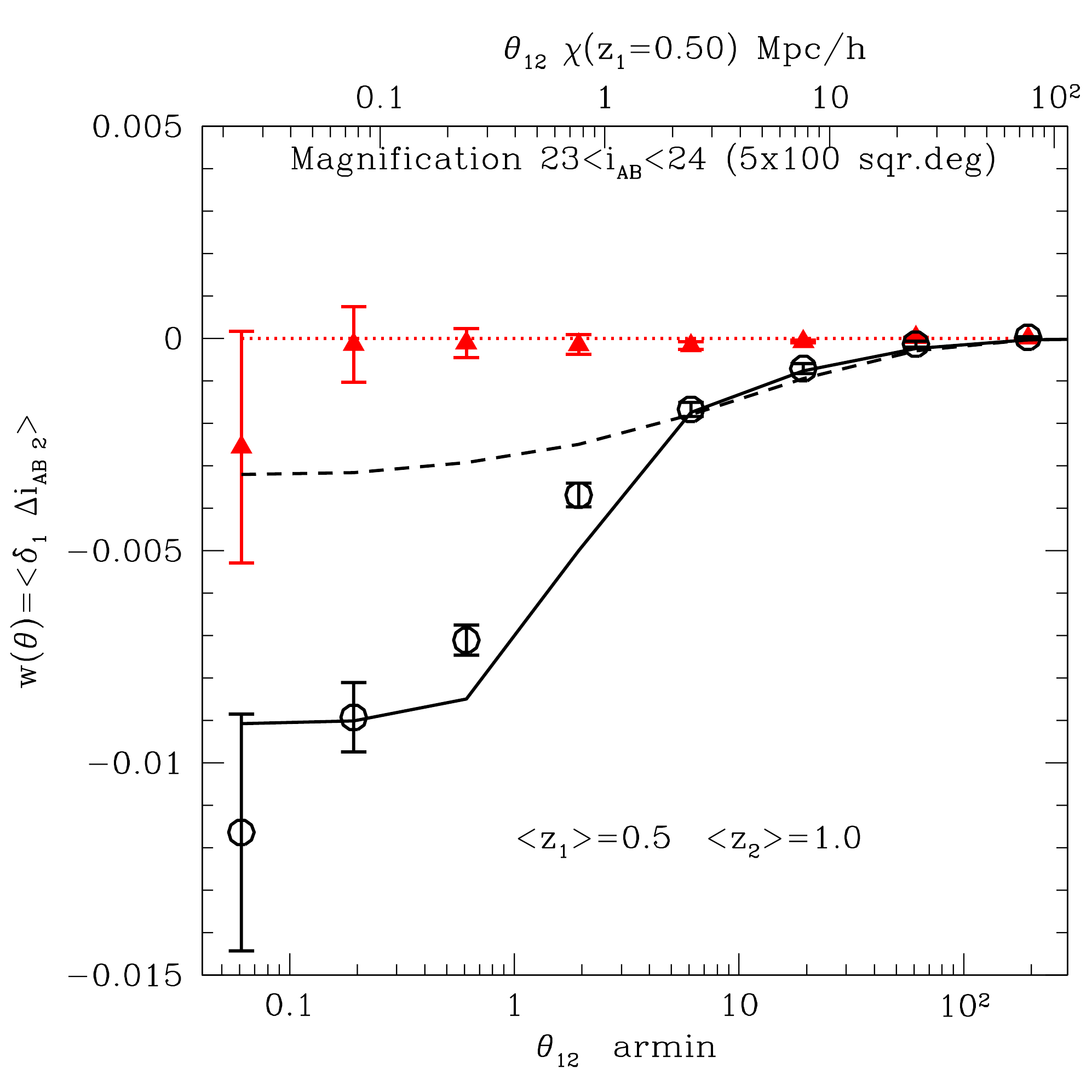}
\caption{Galaxy-magnitude correlation for sources at $z_s=1.0$ and
  lenses at $z_l=0.50$.  Dashed and continuous lines show the linear
and non-linear predictions for DM magnification. Open circles include the
magnification signal in the magnitudes and positions, while closed (red) triangles
only in the positions.}
\label{fig:w2mag}
\end{center}
\end{figure}

\subsubsection{Magnitudes}

In Fig.\ref{fig:w2mag} we show the cross-correlation of foreground densities
with background magnitudes: $\Delta i_{AB} = i_{AB}-<i_{AB}>$. This signal is expected to be proportional
to $< \delta ~  \kappa >$ for the same sample, shown in Fig.\ref{fig:w2gs}:
\beq
< \delta(\theta_1,z_1) ~ \Delta i_{AB} (\theta_2,z_2)   > = \alpha_m <
\delta(\theta_1,z_1)  ~  \delta\mu(\theta_2,z_2)  >
\eeq
where the proportionality
constant $\alpha_m$ is:

\beq
\alpha_m = {2.5 \over{\ln{10}}} \left( -1 + {d<i_{AB}>\over{d
    i^*_{AB}}} \right)
\eeq
where $<i_{AB}>$ is the mean magnitude in the (source) redshift bin
under consideration and $i^*_{AB}$ is the magnitude limit in our
selection. In Fig.\ref{fig:w2mag} we apply magnification to
positions and also to magnitudes, but we select galaxies according
to the true magnitudes, in the range $23<i_{AB}<24$ and $z_s=1.0 \pm 0.05$.
That way  $\alpha_m =-1$ and we can measure directly the
effect of magnification on individual magnitudes. If we use
 magnified (ie observed) magnitudes
to select the MICE galaxies we find $\alpha_m \simeq -0.35$, close to
the values found in  \cite{Menard} for QSO or \cite{Bauer} for LRG galaxies.
In such case we find that the measurement scales with $\alpha_m$, as 
expected. 

 Fig.\ref{fig:w2mag}  shows a good
agreement with the non-linear DM prediction, shown as
a continuous line. The dashed line is for linear theory. These
predictions are the same as the lines in Fig.\ref{fig:w2gs}, 
scaled by $\alpha_m$. The closed
triangles show the measurement without magnification in the
magnitudes, which yields zero correlation, as expected.
 Positions are also altered by magnification for both triangles
and circles, but this produces no signal
when we use magnitudes instead of counts.
Note that the error-bars (for a 5x100 sq.deg. survey) 
are much smaller here than in the
case of the counts in Fig.~\ref{fig:clkg_mag}. This is because the intrinsic
correlation of magnitudes in the sources
are more weakly correlated than the counts
and this reduces the sampling
variance in the cross-correlation error.


\section{Conclusions}   
\label{sec:conclusions}   

In paper I of this series (\cite{paperI}) we presented 
the MICE Grand Challenge Lightcone simulation
(MICE-GC), one of the largest Nbody runs
completed to date.  It contains about 70 billion particles in a 3 $\Gpc$ periodic box. This unique
combination of large volume and fine mass resolution 
allows to resolve the growth of structure form the largest
(linear) cosmological scales down to very small ($\sim$ tens of Kpcs) scales,
well within the non-linear regime.
Therefore, the MICE-GC presents multiple potential applications to study the
clustering and lensing properties of dark-matter and galaxies over a
wide dynamical range, that can be confronted with observations from
wide and deep galaxy surveys. 

Paper II (\cite{paperII})  presented and validated the FoF halo catalogs obtained from the
MICE-GC. These halo catalogs were populated with galaxies using a HOD+SAM technique over one octant of
the sky (\ie 5000 sq.deg) and to the full depth of the lightcone, $z<1.4$.
Extensive properties where also
attributed to this galaxy mock, named {\tt MICECAT v1.0}, whose
clustering was later throughly validated using 2 and 3-point statistics. 

In the last paper of this series (Paper III) 
we have described the construction of high spatial (sub-arcmin) resolution
all-sky lensing maps from the MICE-GC lightcone simulation outputs, 
and discuss their application to model galaxy lensing properties of
the {\tt MICECAT} mock catalog.
These properties, including the convergence, shear, and lensed magnitudes
and positions, have been validated using various lensing
observables in harmonic and real space. In particular, we have
studied auto and cross correlations of both dark-matter and galaxy samples.

Below we provide the main findings from this extensive analysis of
the all-sky lensing maps and derived galaxy lensing mocks:

\begin{itemize}

\item{
Mass resolution effects observed in the angular clustering of
projected dark-matter in the lightcone (see Paper I,
\cite{paperI}), 
are expected to build up in
lensing observables, as they integrate contributions of all matter
shells along the line of sight. This qualitative picture is confirmed
using the lensing maps produced from the MICE simulations (see
Fig\ref{fig:shearmat}): the smallest mass halos resolved in the higher
resolution run, $M_{min} \sim 3\times 10^{11} \Msun$, which have
diameter of $D (M_{min})\sim$ 2 Mpc/h,  are not found 
in previous (lower-mass resolution) runs. As a consequence, and
in the language of the halo model, the 1-halo contribution to the
angular power spectra of the convergence (or shear) field
is substantially suppressed for the low-resolution realizations at the angular
scales subtended by these $M_{min} $ halos, typically few arcminutes
in the sky. This is shown in Fig.\ref{fig:clkmassres} where comparison
of the measured power in MICE-GC with respect to the lower-resolution
MICE-IR run implies that mass-resolution effects are at the 5$\%$ level for $\ell\sim 10^3$
and 20 $\%$ for $\ell\sim 10^4$.}

\item{We find an excess of convergence power for sources at $z=1$ in the MICE-GC
    relative to Halofit \citep{smith03}, at the $20\%$ level for
    $\ell >10^4$, as shown in Fig.\ref{fig:clkmassres}.
This is consistent with a similar analysis from the higher-resolution Millennium
Simulation \citep{hilbert09}.  On the other hand, other 
recent high-resolution simulations seem to find an even larger power excess
at this highly non-linear scales (see \cite{takahashi12}).}

\item{
We have also modelled the lensing
properties of synthetic galaxies using the MICE-GC lightcone
simulation up to $z=1.4$, using the ``onion universe'' approach (see
\cite{fosalba08}) that is equivalent to ray-tracing techniques in the
Born approximation. We have tested the accuracy of our simulation by
comparing the auto-correlation of the convergence and the galaxy
counts-convergence cross-correlation to current numerical
fits. Fig.\ref{fig:clkg} and \ref{fig:w2mag} 
show that for a magnitude limited source
sample at $z_s=1$  there is very good match between MICE and
Halo-model fits from linear scales down to the resolution or pixel scale (i.e, about 1 arcmin.) of the angular maps used for the
analysis. In particular, a simple linear galaxy bias of $b\simeq 1.35$
matches the measured cross-correlation, in agreement with the bias factor
estimated from the foreground galaxy sample auto-correlation at
$z=0.5$ (see Fig.\ref{fig:clgglens})}

\item{
Comparison of the 2-point shear correlation functions measured in MICE 
to halo-model fits shows that our simulation is accurate 
down to $\simeq 2$ arcmin scales. The MICE-GC measurements
underestimate recent theory fits based on high-resolution simulations
\citep{takahashi12} by  $10\%$ on arcmin scales, and overpredicts 
lower resolution halo-model fits \citep{smith03} by a similar amount.}

\item{
In this paper we have also introduced for the first time 
the modeling of the magnification effect on the observed
positions of background galaxies using all-sky lensing maps.
This ``area dilution'' effect induces a measurable anti-correlation between foreground
and distant background galaxy samples. 
The magnification effect in galaxy positions has opposite sign with respect
to the flux magnification by foreground sources in magnitude limited
samples, and therefore the net lensing effect on distant galaxies,
known as {\it magnification bias}
depends on the slope $s$ of the background population source counts,
as shown in Fig.\ref{fig:magbias}.}

\item{
By taking different magnitude limited galaxy samples we have compared
the two qualitatively different contributions to the magnification bias in MICE as
compared to theoretical predictions (in the Born approximation).
As shown in left panels of Fig.\ref{fig:clkg_mag}, simulation results
match well the theory in all the dynamical range (i.e, almost three
decades in angular scale). The oscillations observed in simulation
measurements are believed to be due to the large sample variance that affects
the cross-correlations (ie., auto-correlations contributing to the
noise are comparable to the signal). This is confirmed in the right
panels which show that when building ``sample variance free''
combinations of correlations, the match between simulations and theory
is significantly improved (see text for details).} 

\item{We also present for the first time the modeling of cross-correlations
between foreground counts with (magnified) background magnitudes. 
Our anlaysis shows good agreement between simulation measurements
and theory expectations, as depicted in Fig.\ref{fig:w2mag}.
The error-bars are much smaller here than in the
case of magnification bias above. This is because the intrinsic
correlation of magnitudes in the sources
are more weakly correlated than the counts
and this reduces the sampling
variance in the cross-correlation error.
This and the points above show that our
modeling of magnification in galaxy positions can be safely used down
to arcminute scales.}

 \end{itemize}

We are making a first public data release of the MICE-GC Galaxy mocks, {\tt MICECAT v1.0},
including the lensing properties described in this paper,
through a dedicated webportal, {\texttt http://cosmohub.pic.es}.
This new galaxy mock should serve as a powerful tool 
to model the clustering and lensing observables expected from
upcoming large astronomical surveys in the era of precision cosmology.

\section*{Acknowledgments} 
We would like to thank Christopher Bonnett, Antony Lewis, Guilhem
Lavaux and Carlos Lopez for useful discussions.
We are greatly indebted to Jorge Carretero, Christian Neissner, Davide
Piscia, Santi Serrano and Pau Tellada for their development of the
CosmoHub web-portal.
We acknowledge support from the MareNostrum supercomputer (BSC-CNS, www.bsc.es), through grants 
AECT-2008-1-0009, 2008-2-0011, 2008-3-0010, 2009-1-0009, 2009-2-0013, 2009-3-0013,
and Port d'Informaci\'o Cient\'ifica (PIC, www.pic.es) where the simulations were ran and stored, respectively.
The MICE simulations were implemented using the Gadget-2 code (www.mpa-garching.mpg.de/gadget).
Funding for this project was partially provided by the European
Commission Marie Curie Initial Training Network CosmoComp
(PITN-GA-2009 238356), the Spanish Ministerio de Ciencia e Innovacion (MICINN),
projects 200850I176, AYA-2009-13936, AYA-2012-39620, AYA-2012-39559, 
Consolider-Ingenio CSD2007-00060 and research project SGR-1398 from
Generalitat de Catalunya. 
MC acknowledges support from the Ram{\'o}n y Cajal MICINN program.
  
\bibliography{mnIII.bib}




\end{document}